\input harvmac
%\draftmode
\def \lc {light-cone\ }
\def \ads {$AdS_3 \times S^3$ }
\def \lr {\lref }
\def\bd {{{\del}_-}}
\def \vp {\varphi}
\def \pw {plane-wave\ }

\def \eq#1 {\eqno{(#1)}}

\def \a {\alpha}

\def \b {\beta}
\def \k {\kappa}

\def \om {\omega}

\def \ep {\epsilon}
\def \s {\sigma}

\def \d {\delta}
\def \G {\Gamma}
\def \l {\lambda}
\def \m {\mu}
\def \g {\gamma}
\def \n {\nu}

\def \e#1 {{\rm e}^{#1}}
\def \const {{\rm const }}

\def \vp {\varphi}
\def \ha { { 1\over 2 }}

\def \ov {\over}
\def \tt { \theta}

\def\E {{\cal E}}

\def \bb {{\beta}}

\baselineskip8pt \Title{\vbox {\baselineskip 6pt{\hbox{ }}{\hbox {
}}{\hbox {}}{\hbox{ }} {\hbox{ }}} } {\vbox{\centerline {On
solvable models of type IIB superstring } \vskip4pt \centerline{in
NS-NS and R-R plane wave backgrounds } }} \vskip -20 true pt
\centerline { {J.G. Russo$^{a}$ \footnote {$^*$} {e-mail address:
russo@df.uba.ar } and A.A. Tseytlin$^{b,c}$
\footnote{$^{\star}$}{\baselineskip8pt e-mail address:
tseytlin.1@osu.edu  } \footnote{$^{\dagger}$}{\baselineskip8pt
Also at Lebedev Physics Institute, Moscow.} }}
\medskip \smallskip
\centerline {\it {}$^a$ Departamento de F\'\i sica, Universidad de
Buenos Aires, }
\smallskip
\centerline {\it Ciudad Universitaria and Conicet, Pab. I, 1428
Buenos Aires, Argentina}

\medskip

\centerline {\it {}$^b$ Theoretical Physics Group, Blackett
laboratory }
\smallskip
\centerline {\it Imperial College, London SW7 2BZ, U.K.}

\medskip
%\smallskip\smallskip
\centerline {\it {}$^c$ Smith Laboratory, Ohio State University }
\smallskip

\centerline {\it Columbus OH 43210-1106, USA }
\bigskip
\centerline {\bf Abstract}
\medskip
\baselineskip10pt \noindent
%%%%%%%%%%%%%%%%%%%%%%%%%%%%%%%%%%%%%%%%%%%%%%%%%%%%%%%%%%%%%%%%%%%%
We consider type IIB string in the two plane-wave backgrounds
which may be interpreted as special limits of the $AdS_3 \times
S^3$ supported by the NS-NS or R-R 3-form backgrounds. The NS-NS
plane-wave string model is equivalent to a direct generalization
of the Nappi-Witten model, with its spectrum being similar to that
of strings in constant magnetic field. The R-R model can be solved
in the light-cone gauge, where the Green-Schwarz world-sheet
theory reduces to a system of free fields: 4 massive and 4
massless copies of bosons and fermions. We describe the string
spectra of the two models and study the associated asymptotic
density of states. We also discuss a more general class of exactly
solvable plane-wave models with reduced or completely broken
supersymmetry which are obtained by adding twists in two spatial
2-planes.

\medskip

\Date {February 2002}
\noblackbox
\baselineskip 16pt plus 2pt minus 2pt

\noblackbox \overfullrule=0pt

\def\I{\alpha }
\def\mm { {\rm m}}
\def \te {{\rm t}}

\def \F {{\cal F}}

\def \ii {{k'}} 
\def \H {{\cal H}}

\def \H {{\cal H}}

\def \rf {\refs}
\def \hu { {\hat u}}
\def \hv { {\hat v}}

%%%%%%%%%%%%%%%%%%%%%%%%%%%%%%%%%%%%%%%%%%%%%%%%%%%%%%%%%%%%%%%
%%%%%%%%%%%%%%%%%%%%%%%%%%%%%%%%%%%%%%%%%
%%%%%%%%%%%%%%%%%%%%%%%%%%%%%%%%
\lref \berg {E.~A.~Bergshoeff, R.~Kallosh and T.~Ortin,
``Supersymmetric string waves,'' Phys.\ Rev.\ D {\bf 47}, 5444
(1993) [hep-th/9212030].
}

\lref\hrts{G.~T.~Horowitz and A.~A.~Tseytlin, ``A New class of
exact solutions in string theory,'' Phys.\ Rev.\ D {\bf 51}, 2896
(1995) [hep-th/9409021].
}

\lref\russu{ J.~G.~Russo and L.~Susskind, ``Asymptotic level
density in heterotic string theory and rotating black holes,''
Nucl.\ Phys.\ B {\bf 437}, 611 (1995) [hep-th/9405117].
}

\lref \ant {I.~Antoniadis and N.~A.~Obers, ``Plane gravitational
waves in string theory,'' Nucl.\ Phys.\ B {\bf 423}, 639 (1994)
[hep-th/9403191].
}

\lref \napwi { C.~R.~Nappi and E.~Witten, ``A WZW model based on a
nonsemisimple group,'' Phys.\ Rev.\ Lett.\ {\bf 71}, 3751 (1993)
[hep-th/9310112].
}

\lref \rrt { J.~G.~Russo and A.~A.~Tseytlin, ``Magnetic flux tube
models in superstring theory,'' Nucl.\ Phys.\ B {\bf 461}, 131
(1996) [hep-th/9508068].
}

\lref \tserus{ J.~G.~Russo and A.~A.~Tseytlin, ``Constant magnetic
field in closed string theory: An Exactly solvable model,'' Nucl.\
Phys.\ B {\bf 448}, 293 (1995) [hep-th/9411099].
}

\lref\taka{ T.~Takayanagi and T.~Uesugi, ``Orbifolds as Melvin
geometry,'' JHEP {\bf 0112}, 004 (2001) [hep-th/0110099]. }
\lref\forg{ P.~Forgacs, P.~A.~Horvathy, Z.~Horvath and L.~Palla,
``The Nappi-Witten string in the light-cone gauge,'' Heavy Ion
Phys.\ {\bf 1}, 65 (1995) [hep-th/9503222].
}

\lr\sfets{ K.~Sfetsos and A.~A.~Tseytlin, ``Four-dimensional plane
wave string solutions with coset CFT description,'' Nucl.\ Phys.\
B {\bf 427}, 245 (1994) [hep-th/9404063].
%%CITATION = HEP-TH 9404063;%%
}

\lr\TT{ A.~A.~Tseytlin, ``Exact solutions of closed string
theory,'' Class.\ Quant.\ Grav.\ {\bf 12}, 2365 (1995)
[hep-th/9505052].
}

\lr\super{J.~G.~Russo and A.~A.~Tseytlin, ``Supersymmetric
fluxbrane intersections and closed string tachyons,'' JHEP {\bf
0111}, 065 (2001) [hep-th/0110107].}

\lr \tsst{A.A. Tseytlin, ``Exact string solutions and duality", to
appear in: {\it Proceedings of the 2nd Journ\' ee Cosmologie}, ed.
H. de Vega and N. S\' anchez (World Scientific), hep-th/9407099. }
%%CITATION = HEP-TH 9407099;%%

\lr \kkl {E.~Kiritsis, C.~Kounnas and D.~Lust, ``Superstring
gravitational wave backgrounds with space-time supersymmetry,''
Phys.\ Lett.\ B {\bf 331}, 321 (1994) [hep-th/9404114].
}

\lref \rshet{ J.~G.~Russo and A.~A.~Tseytlin, ``Heterotic strings
in uniform magnetic field,'' Nucl.\ Phys.\ B {\bf 454}, 164 (1995)
[hep-th/9506071].}

\lref \gsw { M.~B.~Green, J.~H.~Schwarz and E.~Witten,
``Superstring Theory. Vol. 1: Introduction,'' {\it Cambridge, UK:
Univ. Pr. ( 1987)}, section 2.3.5. }

\def \sms {$\s$-models\ }

\lref\rrm { R.~R.~Metsaev, ``Type IIB Green Schwarz superstring in
plane wave Ramond Ramond background,'' hep-th/0112044.
}

\lref \lust { E.~Kiritsis, C.~Kounnas and D.~Lust, ``Superstring
gravitational wave backgrounds with space-time supersymmetry,''
Phys.\ Lett.\ B {\bf 331}, 321 (1994) [hep-th/9404114].
%%CITATION = HEP-TH 9404114;%%
}

\lref \blau { M.~Blau, J.~Figueroa-O'Farrill, C.~Hull and
G.~Papadopoulos, ``A new maximally supersymmetric background of
IIB superstring theory,'' hep-th/0110242.
%%CITATION = HEP-TH 0110242;%%
} 

\lr\blap{ M.~Blau, J.~Figueroa-O'Farrill, C.~Hull and
G.~Papadopoulos, ``Penrose limits and maximal supersymmetry,''
hep-th/0201081.
%%CITATION = HEP-TH 0201081;%%
}

\lref \tet{ A.~A.~Tseytlin, ``Closed superstrings in magnetic flux
tube background,'' Nucl.\ Phys.\ Proc.\ Suppl.\ {\bf 49}, 338
(1996) [hep-th/9510041].
}

\lr \mald { D.~Berenstein, J.~Maldacena and H.~Nastase, ``Strings
in flat space and pp waves from N = 4 super Yang Mills,''
hep-th/0202021. }
%%CITATION = HEP-TH 0202021;%%

\lr \hull {C.~M.~Hull, ``Killing Spinors And Exact Plane Wave
Solutions Of Extended Supergravity,'' Phys.\ Rev.\ D {\bf 30}, 334
(1984).
J.~Kowalski-Glikman, ``Vacuum States In Supersymmetric
Kaluza-Klein Theory,'' Phys.\ Lett.\ B {\bf 134}, 194 (1984).
R.~Gueven, ``Plane Waves In Effective Field Theories Of
Superstrings,'' Phys.\ Lett.\ B {\bf 191}, 275 (1987).
%%CITATION = PHLTA,B191,275;%%
}

\lr \mt { R.~R.~Metsaev and A.~A.~Tseytlin, ``Exactly solvable
model of superstring in plane wave Ramond-Ramond background,''
hep-th/0202109.
}

\lr \alos {E.~Alvarez, ``Strings At Finite Temperature,'' Nucl.\
Phys.\ B {\bf 269}, 596 (1986).
%%CITATION = NUPHA,B269,596;%%
E.~Alvarez and M.~A.~Osorio, ``Superstrings At Finite
Temperature,'' Phys.\ Rev.\ D {\bf 36}, 1175 (1987).
%%CITATION = PHRVA,D36,1175;%%
} \lr\kk{ E.~Kiritsis and C.~Kounnas, ``String Propagation In
Gravitational Wave Backgrounds,'' Phys.\ Lett.\ B {\bf 320}, 264
(1994) [Addendum-ibid.\ B {\bf 325}, 536 (1994)] [hep-th/9310202].
%%CITATION = HEP-TH 9310202;%%
}
%%%%%%%%%%%%%%%%%%%%%%%%%%%%%%%%%%%%%%%%%%%%%%%%%%%%%%%%%%%%%%%%%%

\lr \all{ D.~Amati and C.~Klimcik, ``Strings In A Shock Wave
Background And Generation Of Curved Geometry From Flat Space
String Theory,'' Phys.\ Lett.\ B {\bf 210}, 92 (1988).
%%CITATION = PHLTA,B210,92;%%
G.~T.~Horowitz and A.~R.~Steif, ``Space-Time Singularities In
String Theory,'' Phys.\ Rev.\ Lett.\ {\bf 64}, 260 (1990).
``Strings In Strong Gravitational Fields,'' Phys.\ Rev.\ D {\bf
42}, 1950 (1990).
%%CITATION = PHRVA,D42,1950;%%
G. Horowitz, in: {\it Proceedings of Strings '90}, College
Station, Texas, March 1990 (World Scientific,1991). H.~J.~de Vega
and N.~Sanchez, ``Space-Time Singularities In String Theory And
String Propagation Through Gravitational Shock Waves,'' Phys.\
Rev.\ Lett.\ {\bf 65}, 1517 (1990).
}

\lr\mall{ R.~E.~Rudd, ``Compactification Propagation,'' Nucl.\
Phys.\ B {\bf 352}, 489 (1991).
C.~Duval, G.~W.~Gibbons and P.~Horvathy, ``Celestial Mechanics,
Conformal Structures And Gravitational Waves,'' Phys.\ Rev.\ D
{\bf 43}, 3907 (1991).
%%CITATION = PHRVA,D43,3907;%%
A.A. Tseytlin, ``String vacuum backgrounds with covariantly
constant null Killing vector and 2-d quantum gravity,'' Nucl.\
Phys.\ B {\bf 390}, 153 (1993) [hep-th/9209023].
%%CITATION = HEP-TH 9209023;%%
}

\lr\klts{ C.~Klimcik and A.~A.~Tseytlin, ``Duality invariant class
of exact string backgrounds,'' Phys.\ Lett.\ B {\bf 323}, 305
(1994) [hep-th/9311012].
%%CITATION = HEP-TH 9311012;%%
}

\lr\jofre{ O.~Jofre and C.~Nunez, ``Strings In Plane Wave
Backgrounds Revisited,'' Phys.\ Rev.\ D {\bf 50}, 5232 (1994)
[hep-th/9311187].
%%CITATION = HEP-TH 9311187;%%
}

\lr \schwarz{ J.~H.~Schwarz, ``Covariant Field Equations Of Chiral
N=2 D = 10 Supergravity,'' Nucl.\ Phys.\ B {\bf 226}, 269 (1983).
%%CITATION = NUPHA,B226,269;%%
}

\lr\papa{ M.~Blau, J.~Figueroa-O'Farrill and G.~Papadopoulos,
``Penrose limits, supergravity and brane dynamics,''
hep-th/0202111.
} \lr \itz{N. Itzhaki, I.R. Klebanov and S. Mukhi, ``PP Wave Limit
and Enhanced Supersymmetry in Gauge Theories'', hep-th/0202153.}

\lr \oog{ J. Gomis and H. Ooguri, ``Penrose Limit of N=1 Gauge
Theories'', hep-th/0202157.}

\lr\guv{R.~Gueven, ``Plane wave limits and T-duality,'' Phys.\
Lett.\ B {\bf 482}, 255 (2000) [hep-th/0005061].
}

\lr \olive{K.~Sfetsos, ``Gauging a nonsemisimple WZW model,''
Phys.\ Lett.\ B {\bf 324}, 335 (1994) [hep-th/9311010].
%%CITATION = HEP-TH 9311010;%%
D.~I.~Olive, E.~Rabinovici and A.~Schwimmer, ``A Class of string
backgrounds as a semiclassical limit of WZW models,'' Phys.\
Lett.\ B {\bf 321}, 361 (1994) [hep-th/9311081].
}

\lref \mplt { A.~A.~Tseytlin, ``Extreme dyonic black holes in
string theory,'' Mod.\ Phys.\ Lett.\ A {\bf 11}, 689 (1996)
[hep-th/9601177].
}

\lr\stan{S.~Stanciu and A.~A.~Tseytlin, ``D-branes in curved
spacetime: Nappi-Witten background,'' JHEP {\bf 9806}, 010 (1998)
[hep-th/9805006].
%%CITATION = HEP-TH 9805006;%%
J.~M.~Figueroa-O'Farrill and S.~Stanciu, ``More D-branes in the
Nappi-Witten background,'' JHEP {\bf 0001}, 024 (2000)
[hep-th/9909164].
} \lr\peet{ A.~W.~Peet, ``TASI lectures on black holes in string
theory,'' hep-th/0008241.
}

\lr\oogm{ J.~Maldacena and H.~Ooguri, ``Strings in AdS(3) and
SL(2,R) WZW model. I,'' J.\ Math.\ Phys.\ {\bf 42}, 2929 (2001)
[hep-th/0001053].
``Strings in AdS(3) and the SL(2,R) WZW model. III: Correlation
functions,'' hep-th/0111180.
}

\lr \mye{ J.~C.~Breckenridge, R.~C.~Myers, A.~W.~Peet and C.~Vafa,
``D-branes and spinning black holes,'' Phys.\ Lett.\ B {\bf 391},
93 (1997) [hep-th/9602065].
}

\lr \gnw{ N.~Mohammedi, ``On bosonic and supersymmetric current
algebras for nonsemisimple groups,'' Phys.\ Lett.\ B {\bf 325},
371 (1994) [hep-th/9312182].
K.~Sfetsos, ``Gauged WZW models and nonAbelian duality,'' Phys.\
Rev.\ D {\bf 50}, 2784 (1994) [hep-th/9402031].
J.~M.~Figueroa-O'Farrill and S.~Stanciu, ``Nonsemisimple Sugawara
constructions,'' Phys.\ Lett.\ B {\bf 327}, 40 (1994)
[hep-th/9402035].
A.~A.~Kehagias and P.~A.~Meessen, ``Exact string background from a
WZW model based on the Heisenberg group,'' Phys.\ Lett.\ B {\bf
331}, 77 (1994) [hep-th/9403041].
A.~A.~Kehagias, ``All WZW models in D <= 5,'' hep-th/9406136.
}

\newsec{Introduction}
%%%%%%%%%%%%%%%%%%%%%%%%%%%%%
Plane-wave backgrounds are exact solutions of string theory (see,
e.g., \refs{\all,\mall,\klts,\jofre,\hrts}, and \TT\ for a
review). Certain NS-NS \pw backgrounds with non-zero 3-form field
correspond to (gauged) WZW models \refs{\napwi,\kk,\olive,\sfets,
\lust,\ant}. More generally, a class of gravitational \pw
backgrounds lead to exactly solvable (in terms of free oscillators
in \lc gauge) class of non-compact curved-space (super)string
models, where one can explicitly find the string spectrum and
compute simplest correlation functions in much the same way as in
flat space. As was recently realised \refs{\rrm,\mald,\mt}, this
solvability property applies to string models corresponding not
only to the NS-NS but also to certain R-R \rf{\blau,\blap} \pw
backgrounds.

The simplest gravitational \pw metric $$ds^2 = dudv - K(u,x) du^2
+ dx_i dx_i$$ \ solves the vacuum Einstein (and, in fact, full
string-theory) equations provided $ \del_i \del^i K =0$ \
($i=1,...,D-2$). Particular solutions are $K_1= w_{ij} (u) x_i
x_j$ with $ w_{ii}=0$ or
$K_2 = { Q(u) \ov |x|^{D-4} }$ (for $x\not=0$). Fixing the \lc
gauge, $u= 2\a' p^u \tau$, we thus get the following effective
Lagrangian for the transverse coordinates:
$$L = \del_+ x_i \del_- x_i - V(\tau,x)\ , \ \ \ \ \ \ \ \ \
V = (\a' p^u)^2 K(2\a'p^u\tau,x) \ . $$ This gives a quadratic,
and thus exactly solvable, string theory in the case of $K_1$ (but
not in the case of $K_2$). For example, one may consider $K= w(u)
( x^2_1 - x^2_2)$, solve explicitly the linear classical string
equations, and then perform the canonical quantization (cf.
\refs{\all,\jofre}).

The trace-free condition on the matrix $w_{ij}$ following from the
Ricci-flatness of the metric implies that the corresponding
quadratic potential $V$ is not positive definite in at least one
of the directions.\foot{ As a result, the corresponding \lc string
spectrum (i.e. spectrum of $H= - \ha p^v$) is continuous. For
constant $w$ this is clear, for example, from the form of the
target-space Laplace operator written in momentum representation
in $(u,v)$ directions: $ \nabla^2 \to \del_i^2 - (p^u)^2 w_{ij}
x_i x_j - p^u p^v $.} It is possible to obtain models with
non-negative $V$ by adding extra fields with constant null field
strengths which produce non-vanishing $(uu)$-component of the
stress tensor and thus change the condition $w_{ii}=0$ into
$w_{ii} = F^2 > 0$. As a result, the (non-trivial part of)
corresponding string spectrum becomes discrete. For example, the
Nappi-Witten (NW) \napwi\ background which has a non-zero NS-NS
3-form field $H_{u12} = -2f $ as well as the \pw metric with $K=
f^2 (x_1^2 + x_2^2)$.

The same \pw metric may be supported by different combinations of
NS-NS or R-R backgrounds, leading to different string models with
different properties. For example, the metric with $K= f^2 (x^2_1
+ ...+x^2_8)$ can be supported by the 5-form background leading to
the string model \refs{\rrm,\mald,\mt} with maximal supersymmetry
\blau, or by a combination of NS-NS and R-R 3-form fluxes giving a
model with 1/2 of maximal supersymmetry.

Here we shall study in detail two superstring models which
represent fundamental strings propagating in the the same \pw
metric supported by two different -- ``S-dual'' -- backgrounds:
the NS-NS 3-form field in one case, and the R-R 3-form field in
another. These string models are of interest, in particular,
because they are related by a special limit to the $AdS_3 \times
S^3$ models with NS-NS and R-R 3-form fluxes, respectively. As was
discussed in \sfets, starting with a string model in generic NS-NS
background and applying a boost combined with a rescaling of
coordinates and $\a' \to 0$ limit (i.e. performing a string-theory
generalization of the Penrose limit \TT) one finishes with a
particular \pw string model associated with the original
background. In the case of the NS-NS $AdS_3 \times S^3$ model,
i.e. $SL(2,R) \times SU(2)$ WZW theory, this limit produces
\refs{\olive,\sfets} a \pw model which is a direct generalization
of the NW model, i.e. the WZW theory for the group $E^c_2$ \napwi.
The direct \lc gauge solution of the corresponding superstring
model was previously given in \refs{\tserus,\rshet}.

The same limit can be applied to string models in R-R
backgrounds.\foot{ For a detailed discussion of the Penrose limit
at the supergravity level in the general NS-NS/R-R background see
\refs{\guv,\blap,\papa}. See also refs. \refs{\itz,\oog} extending
\mald\ to $AdS_5 \times M^5$ cases.} The \pw model obtained as the
limit of the $AdS_3 \times S^3$ R-R model is very similar to the
R-R 5-form \pw model \rrm\ which is the Penrose limit
\refs{\blap,\mald} of the $AdS_5 \times S^5$ string model.
Following \refs{\rrm,\mt} here we shall determine the
corresponding \lc gauge form of the superstring action and find
the string spectrum (similar final result for the spectrum
was given earlier in Appendix C of \mald).

In section 2 we shall discuss the actions of the two equivalent
NS-NS superstring models (which are essentially the 2-parameter
generalization of the NW model), one having non-diagonal in
$(u,x_i)$ target space metric but chiral 2-d structure (with the
interaction term in the \lc gauge action having the form of
coupling to a constant magnetic field background) while another
having the ``standard'' \pw metric. Their relation via the Penrose
limit to the $AdS_3 \times S^3$ NS-NS model will be reviewed in
section 3. In section 4 we shall present the \lc spectra of the
string states of these two NS-NS models.

The R-R model will be considered in section 5. We shall first
determine the structure of the corresponding quadratic
Green-Schwarz \lc gauge action, and then solve the classical
equations and quantize the theory in a similar way as in
\refs{\rrm,\mt}. We shall also demonstrate that the zero-mode
(supergravity) part of the spectrum coincides with the
corresponding part of spectrum of the NS-NS model, as expected
from S-duality symmetry of type IIB supergravity.

In section 6 we shall study the asymptotic (large \lc energy) form
of the density of states $d_\E$ of the string spectra. We shall
argue that in both the R-R and the NS-NS cases the leading
behaviour of $d_\E$ is the same as in flat space. This is the
expected conclusion, since the dependence of the string spectrum
on the curvature scale should become negligible at large level
numbers.

In section 7 we shall discuss generalizations of the NS-NS models
to less supersymmetric cases by adding Melvin model type twists in
the two spatial 2-planes. One particular case we shall consider
corresponds to compact \lc directions, were supersymmetry is
broken unless the two curvature parameters (as well as the two
twist parameters) are chosen to be equal. The resulting string
spectra have rather non-trivial form and may be of interest in
other contexts as well. We shall note also that compactifying the
\lc directions in the R-R model reduces the number of
supersymmetries by half.

Section 8 will contain some concluding remarks.

%%%%%%%%%%%%%%%%%%%%%%%%%%%%%%
\newsec{Generalized plane wave models with NS-NS 3-form
background}
%%%%%%%%%%%%%%%%%%%%%%%%%%%%%

Here we shall consider two equivalent conformal models, the
bosonic parts of which are special cases of (see \TT)\foot{ We
shall use Minkowski world-sheet coordinates with $\s^\pm =\tau\pm
\s $, and $\del_\pm = \ha (\del_0 \pm \del_1$). The string action
is $S= {1 \ov \pi \a' } \int d^2 \s \ L$. The space-time \lc
coordinates are $u=y-t, \ v= y +t $. } \eqn\stre{ L = \del_+ u \bd
v + K(u,x)\del_+u \bd u + 2A_i(u,x) \del_+ u \bd x^i + 2\bar
A_i(u,x) \del_+ x^i \bd u + \del_+ x^i \bd x^i \ . }

%%%%%%%%%%%%%%%%%%%%%%%%%%%%%%%%%%
\subsec{Bosonic actions}
%%%%%%%%%%%%%%%%%%%%%%%%%%%%%%%%%%%

When $\bar A_i =0$ this model and its direct type II superstring
generalization \hrts\ are conformally invariant provided
\eqn\sses{- \ha \del^2 K + \del^i\del_u A_i =0 \ , \ \ \ \ \ \ \ \
\del^j \F_{ij} =0 \ . } A particular case \eqn\trii{ K=0\ , \ \ \
\ \ \ \ A_i=-\ha \F_{ij} x^j\ , \ \ \ \ \F_{ij} =\const \ , } i.e.
\eqn\sto{ L = \del_+u \bd v + \F_{ij} x^i \del_+u \bd x^j +
\del_+x^i \bd x^i \ , \ \ \ \ \ \ \ \ \ \ i,j=1,...,n\ , } can be
interpreted \hrts\ as boosted products of group spaces, or,
equivalently, as spaces corresponding to WZW models for
non-semisimple groups (see \gnw). The NW model \napwi\ is the
simplest ($D=4$, $n=2$) example with $A_i=- f \epsilon_{ij} x^j$.
\foot{These \sms (or corresponding non-semisimple WZW models) can
be obtained \refs{\olive,\sfets} by singular boosts and rescalings
of levels from semisimple WZW models based on direct products of
$SL(2,R)_{-k}, \ SU(2)_k$ and $R$ factors. The models \stre,\trii,
can be formally related by $O(d,d;R)$ duality to the flat space
model \hrts\ in the same way as for the $D=4$ NW model
\refs{\kk,\klts}.}

Here the NS-NS 2-form has the non-trivial components $B_{uj} = \ha
\F_{ij} x^i$, so that $H_{uij} = -\F_{ij}$. The corresponding
Lorentz connection with torsion ($\omega_{\pm} = \om \pm \ha H $)
has the following non-zero components \hrts\ \eqn\nonzer{\omega_{-
\hat u \hat i} = \ha \del_i K du + \F_{ij} dx^j = \F_{ij} dx^j\ ,
\ \ \ \omega_{+ \hat i\hat j} = - \F_{ij} du \ , \ \ \
\omega_{+\hat u \hat i} = (\ha \del_i K - \del_u A_i) du =0 \ . }
Another conformal model we shall consider is\foot{To check
normalizations note that $R_{uu} = - \ha \del^2 K = w_{ii} , \ \ {
1 \ov 4} H_{uij} H_{uij} = { 1 \ov 4} \F_{ij} \F_{ij}. $}
\eqn\stro{ L = \del_+ u \bd v - w_{ij} x^i x^j \del_+u \bd u + \ha
\F_{ij} x^i ( \del_+u \bd x^j - \bd u \del_+x^j) + \del_+x^i \bd
x^i \ , }
$$
w_{ii} = { 1 \ov 4 } \F_{ij} \F^{ij}\ . $$ Below we shall be
interested in the case when \eqn\cass{ w_{ij} = f^2 \delta_{ij}\ ,
\ \ \ \ \ \ \ i,j=1,..., n \ , } where $n$ is the dimension of
$x^i$ space. $n=2$ corresponds again to the $D=4$ NW model \napwi:
$ \F_{ij} = 2f\ep_{ij}$.

The two models \sto\ and \stro\ have the same 3-form field and
their metrics are related by a $u$-dependent coordinate
transformation: $u$-dependent rotations in the independent $x^i$
planes where $\F_{ij}$ is non-zero (see next section and for $n=2$
also \refs{\ant,\hrts,\forg}). When $u$ is non-compact, the two
models are thus equivalent. These models have the R-R analogs
where the NS-NS 3-form is replaced by the R-R one.

The models \sto\ or \stro\ are exactly solvable since in the \lc
gauge $u\sim p^u \tau$ the transverse theory becomes gaussian. The
same happens also in the fermionic sector. The solution of the
bosonic $n=2$ model (i.e. the NW model) interpreted as a WZW
theory was discussed in \kk. In the general case of compact $u $
(periodic $u+v$ coordinate) it was solved explicitly in \tserus.
In the case of non-compact $u$ its solution in the \lc gauge was
also discussed in \forg. The supersymmetric case of NW model was
discussed from WZW CFT point of view in \lust\ and was solved
explicitly in the \lc gauge in \rshet. Below we will generalize
the string spectra found in \refs{\tserus,\rshet} to the $n=4$
(i.e. two 2-plane) case, i.e. where \eqn\viv{ \F_{kl} = 2f_1
\ep_{kl} \ \ (k,l=1,2) \ , \ \ \ \ \ \ \ \ \ \ \F_{k'l'} = 2f_2
\ep_{k'l'} \ \ (k',l'=3,4) \ , } so that \stro\ becomes (we ignore
free decoupled coordinates)
$$ L = \del_+u \bd v -
( f^2_1x_k^2 + f_2^2 x_\ii^2) \del_+u \bd u + \del_+x^k \bd x^k +
\del_+x^\ii \bd x^\ii $$ \eqn\zs{ + \ f_1 \ep_{kl} x^k ( \del_+u
\bd x^l - \bd u \del_+x^l) + f_2 \ep_{k'l'} x^{k'} ( \del_+u \bd
x^{l'} - \bd u \del_+x^{l'}) \ . } The $n=4$ model we shall be
interested in is related by a limit to \ads NS-NS model \ where
$\F_{ij}$ is the $4\times 4$ self-dual constant matrix, i.e.
\eqn\eqq{ f_1 =f_2 =f \ , } so that \zs\ takes the following
explicit form
$$ L = \del_+u \bd v -
f^2 (x_k^2 + x_\ii^2) \del_+u \bd u + \del_+x^k \bd x^k +
\del_+x^\ii \bd x^\ii $$ \eqn\zro{ + \ f \big[ \ep_{ij} x^i (
\del_+u \bd x^j - \bd u \del_+x^j) + \ep_{i'j'} x^{i'} ( \del_+u
\bd x^{j'} - \bd u \del_+x^{j'})] \ . }

%%%%%%%%%%%%%%%%%%%%%%%%%%%%%%%%%%%
\subsec{Fermionic actions}
%%%%%%%%%%%%%%%%%%%%%%%%%%%%%%%%%%%%%%%%

Let us now discuss the fermionic terms in the corresponding sigma
model action. The fermionic part of (1,1) supersymmetric model
corresponding to \sto\ is \rf{\hrts,\tserus,\rshet}\foot{ The
hatted indices are tangent-space indices, with only $\psi^{\hat
v}$ being different from $\psi^v$.} \eqn\supe{ L_F=\psi^\hu_L
\del_- \psi^\hv_L + \psi^i_L \del_- \psi^i_L + \F_{ij} \del_- x^j
\psi^\hu_L\psi^i_L + \psi^\hu_R\del_+ \psi^\hv_R + \psi^i_R \del_+
\psi^i_R - \F_{ij} \del_+ u \psi^i_R\psi^j_R \ . } In the \lc
gauge \eqn\lcc{ u=2 \a' p^u \tau, \ \ \ \ \ \ \ \ \ \ \ \
\psi^u_{L,R}=0 } it becomes \eqn\upe{ L_F= \psi^i_L \del_-
\psi^i_L + \psi^i_R( \d_{ij} \del_+ - \a' p^u \F_{ij} ) \psi^{
j}_R \ . } The coupling to $\F_{ij}$ here is similar to a coupling
of a superparticle to a magnetic field.
% as required in fact by the 2-d supersymmetry of the model.

The analogous \lc gauge action corresponding to the model \stro\
has ``magnetic'' couplings in the both left and right sectors,
with the opposite signs (reflecting opposite $H_3$ contributions)
\eqn\upeq{ L_F= \psi^i_L ( \d_{ij} \del_- +\ha \a' p^u \F_{ij} )
\psi^i_L + \psi^i_R ( \d_{ij} \del_+ - \ha \a' p^u \F_{ij} )
\psi^i_R \ . } Like the bosonic actions, the two fermionic actions
\upe\ and \upeq\ are related by coordinate rotations in each
eigen-plane of $\F_{ij}$ with phase proportional to $f_k u$.

The corresponding fermionic actions in the Green-Schwarz approach
are also quadratic if one chooses the \lc gauge \eqn\gau{ u = 2
\a' p^u \tau\ , \ \ \ \ \ \ \G^{u} \theta^I=0 \ . } Then the only
coupling to the background is through the spinor covariant
derivatives ($d + { 1 \ov 4} \omega_{\pm \hat m\hat n} \G^{\hat
m\hat n}$). In the case of the background corresponding to \sto\
\eqn\hav{ D_+ \theta = d \theta - { 1 \ov 4} \F_{ij} \Gamma^{ij}
du\ \theta \ , \ \ \ \ D_- \theta = d \theta + { 1 \ov 4} \F_{ij}
\Gamma^{ui} dx^j \theta \ , } so that in the \lc gauge one is left
with the quadratic action containing one extra term compared to
the flat space GS action (cf. \upe) \eqn\fef{ L_F= i S_L \del_-
S_L + i S_R ( \del_+ - {1 \ov 4}\a' p^u \F_{ij} \gamma^{ij} ) S_R
\ . } Here $S_L,S_R$ are the $SO(8)$ spinors. This action has the
same structure as \upe , found in the NSR approach.

The \lc GS action corresponding to the model \stro\ has, like
\upeq, both left and right fermions coupled to $p^u \F_{ij}$ (here
$\om_{\pm i j} = \mp \ha \F_{ij} du$) \eqn\oup{ L_F= i S_L (
\del_- + {1 \ov 8} \a' p^u \F_{ij} \g^{ij} ) S_L + i S_R ( \del_+
- {1 \ov 8} \a' p^u \F_{ij} \g^{ij} ) S_R \ . } Again, the two
actions \fef\ and \oup\ are related by a local Lorentz rotation of
the space-time fermions.

%%%%%%%%%%%%%%%%%%%%%%%%%%%%%
\subsec{Supersymmetry}
%%%%%%%%%%%%%%%%%%%%%%%%%%%%%

%%%%%%%%%%%%%%%%%%%%%%%%%%%%%%%%%%%%%%%%%%%%%%%%%%%%%%%%%%%%%%%%%%%%%%%%%
Let us comment on supersymmetry of these models.
As is well known, pure \pw background (with $H_{\mu\nu\rho}=0$)
preserves 1/2 of maximal space-time supersymmetry (see, e.g.,
\refs{\hull,\berg}).\foot{ The existence of a global supersymmetry
(i.e. the analog of the flat space invariance of the GS action
under $\theta \to \theta + \epsilon, \ e^{\hat \mu } \to e^{\hat
\mu } + i\bar \epsilon^I \Gamma^{\hat \mu} D\theta^I, \ \ e^{\hat
\mu }=e^{\hat \mu }_\nu dx^\nu $) is the same as the Killing
spinor condition $D\epsilon=0$, i.e. $(\del_u - \ha \sum_i f_{i}
x_i \Gamma^{ iu}) \epsilon^I=0 , \ \del_i \epsilon^I=0$. It has
solution $\epsilon= \epsilon_0$=const, where $\Gamma^{
u}\epsilon_0 =0$, i.e. we get 1/2 of maximal supersymmetries
preserved.} In the presence of a $ H_{uij}=-\F_{ij}$, the form of
the \lc Green-Schwarz action \oup\ corresponding to \stro\ (or
\fef\ corresponding to \sto) implies that the Killing spinor
equations are $ ( \del_u \mp { 1 \ov 8} \F_{ij} \Gamma^{ij} )
\theta_{L,R} =0 , \ \ \del_i \theta_{L,R} =0, $ with $\Gamma^u
\theta_{L,R} =0 $. When $u$ is non-compact as in the case we are
interested in here, one can always solve these equations for any
$\F_{ij}$ and find a consistent Killing spinor. Thus the amount of
supersymmetry is the same (i.e. 1/2) as in the absence of $H_3$
background (see \refs{\hrts,\berg}).

If $u$ is a compact coordinate (as in the ``magnetic'' models of
\refs{\tserus,\rshet}, see also section 7), then the models
\sto,\fef\ and \stro,\oup\ are no longer equivalent (for generic
values of the background field parameters) and have different
amounts of space-time supersymmetry. In the case of the ``chiral''
model \sto,\fef\ considered in \refs{\hrts,\tserus,\rshet}
constant shifts of $SO(8)$ spinor $S_L$ are always a symmetry,
while there is no symmetry in $S_R$ sector for generic $\F_{ij}$.
This model will thus have additional supersymmetry in the $S_R$
sector, i.e. the two-parameter generalization of the constant
magnetic model of \tserus\ will preserve 8+4 real supersymmetries
in the $f_1=f_2$ case (analogous mechanism is at work in the
supersymmetric magnetic models of \super).

In the case of the model \stro,\oup\ with compact $u$ all
supersymmetry will be again broken if $\F_{ij}$ is generic.
However, fractions of supersymmetry may be preserved for special
choices of $\F_{ij}$. For example, for a self-dual $\F_{ij}$, i.e.
in the case of $f_1=f_2$, there will be constant Killing spinors
in both $S_L$ and $S_R$ sectors of \oup\ provided they satisfy $
(\gamma_{12} + \gamma_{34}) S_{L.R}=0, \ $ i.e. $\gamma_{1234}
S_{L,R} =S_{L,R}. $ Thus \oup\ will preserve 1/4 of the maximal
number of supersymmetries, i.e. will have 8 real supercharges.

%%%%%%%%%%%%%%%%%%%%%%%%%%%
\newsec{``Plane-wave" limit of $AdS_3\times S^3$ string sigma model }

The model \zs\ arises as a limit of \ads with 3-form flux or
$SL(2,R) \times SU(2)$ WZW model following the limiting procedure
described in \refs{\sfets,\blap}. This limit can be taken directly
in the string sigma model action (which remains invariant under
the corresponding rescaling of coordinates and parameters) or
separately in the metric and $B_{\m\n}$. Let us start with the
following parametrization of the metric of \ads \eqn\uno{ ds^2 =
f^2 \big[ - (1+r_1^2)dt^2 + {dr_1^2\over 1+ r_1^2} + r_1^2
d\varphi_1 ^2 + (1-r_2^2)d\psi^2 + {dr_2^2\over 1- r_2^2} + r_2^2
d \varphi_2^2 \big] \ . } Let us rescale $r_1, r_2$ by $\epsilon
$, $f\to f/\epsilon $, set \eqn\jeep{ t=u\ ,\ \ \ \ \ \psi = u +
\ha \epsilon^2 v\ , } and consider the limit $\epsilon\to 0$.
\foot{The rescaling of the size of the space $f^2=R^2/\a'$ is
equivalent to $\a' \to 0$, explaining why the resulting background
solves string equations to all orders in $\a'$.} This gives us the
plane-wave metric: \eqn\ftres{ ds^2 = dudv - f^2 x_i^2 du^2+ dx^i
dx^i \ , \ \ \ \ \ \ \ \ \ \ \ \ i=1,...,4 \ , } where we have
introduced Cartesian coordinates $x^i$, \ \ $$ dx^i dx^i= dr_1^2
+r_1^2\varphi_1 ^2+dr_2^2+ r_2^2 d \varphi_2^2 \ , $$ and made a
finite rescaling of $v$ and $x^i$ by powers of $f$. The components
of $B_{\mu\nu}$ are found by the same limiting procedure, with the
overall rescaling compensated by rescaling of $f$.

The final result is the \pw sigma model action \zro.
Note that if $t$ is non-compact, both $u$ and $v$ are non-compact 
in the limit $\epsilon \to 0$: the period of $v$ which is equal to 
$8\pi/\epsilon^2$ becomes infinite.

 For
generality, in what follows we shall keep $f_1,f_2$ parameters
arbitrary. The case of $f_1=f_2$ corresponding to the limit of
\ads has $SO(4)$ invariant metric, but this symmetry is in any
case broken down to $SO(2) \times SO(2)$ by the antisymmetric
tensor terms.

The model \zs\ is equivalent to \eqn\fsto{ L = \del_+u \bd v + 2
f_1 \ep_{kl} x^k \del_+u \bd x^l + 2 f_2 \ep_{k'l'} x^{k'} \del_+u
\bd x^{l'}+ \del_+x^{k} \bd x^{k}+ \del_+x^{k'} \bd x^{k'} \ . }
The lagrangian \fsto\ is a particular case of \sto \ with
block-diagonal $4 \times 4$ matrix ${\cal F}_{ij} $. In polar
coordinates in the two planes it takes the form
$$ L =
\del_+u \bd v + 2 f_1 r_1^2 \del_+u \bd \varphi_1 + 2 f_2 r_2^2
\del_+ u \bd \varphi_2 $$ \eqn\stoi{ +\ \del_+r_1 \bd
r_1+r_1^2\del_+\varphi_1\bd\varphi_1 +\del_+r_2 \bd
r_2+r_2^2\del_+\varphi_2\bd\varphi_2\ . } The model \zs\
(corresponding to the metric \ftres\ for $f_1=f_2$) is related to
\stoi\ by the sigma model field redefinition \foot{From the target
space point of view this is a coordinate transformation combined
with the gauge transformation $B_{\mu\nu}\to
B_{\mu\nu}+2\del_{[\mu}\Lambda_{\nu ]}$, \ $\Lambda_i=u f
\epsilon_{ij} x^j$. Note that this coordinate transformation is
globally defined as long as $u$ is non-compact. If $u$ is compact,
a similar redefinition in the two planes corresponds to twists
considered in connection with ``rotating'' NS5-NS1 system (related
to rotating 5-d black holes \mye) in \mplt\ and also (in the case
when $u$ is replaced by a space-like coordinate) in connection
with the Melvin-type ``magnetic'' models in \super . } \eqn\hhoo{
\varphi_1=\varphi_1' - f_1 u\ ,\ \ \ \ \ \ \ \
\varphi_2=\varphi'_2 - f_2 u\ . \ } Indeed, it transforms \stoi\
into
$$ L =
\del_+u \bd v - (f_1^2r_1^2 +f_2^2r_2^2)\del_+u \bd u + \del_+r_1
\bd r_1+r_1^2\del_+\varphi_1'\bd\varphi_1' +\del_+r_2 \bd
r_2+r_2^2\del_+ \varphi_2'\bd\varphi_2'
$$ \eqn\cstro{
+\ f_1 r_1^2 ( \del_+u\bd \varphi_1' -\bd u\del_+\varphi_1') +f_2
r_2^2 ( \del_+u\bd \varphi_2' -\bd u\del_+\varphi_2') \ , } which
is equivalent to \zs.

The fermions are included in a straightforward way, according to
\upeq\ for \zs\ and \upe\ for \fsto.

%%%%%%%%%%%%%%%%%%%%%%%%%%%%%%%%
\newsec{Solution of the NS-NS plane wave superstring model}
%%%%%%%%%%%%%%%%%%%%%%%%%%%%%%%%
Below we shall first discuss the spectrum of the two models \fsto\
and \zs. We shall only consider the case of $y=\ha (u+v)$ being
non-compact when they are equivalent (cf. \hhoo) and are related
to the \pw limit of the \ads theory. In the \lc gauge they are
simply direct products of the NW model discussed in
\refs{\tserus,\forg}, so the spectrum is readily obtained from the
previous results.

The conformal model \fsto\ can be solved by a simple extension of
the single 2-plane case ($f_2=0$) studied in \tserus\ for the
bosonic case, and \rshet\ for the supersymmetric case. Let us set
(note that $\te$ is different from $t$ in \jeep) \eqn\ett{ u= y -
\te \ , \ \ \ \ \ \ \ v = y + \te \ , } In \refs{\tserus,\rshet },
the coordinate $y$ was compact describing a circle of radius $R$.
In the limit $R\to \infty $ we are interested in here the charges
$Q_{L}$, $Q_R$ of \rshet\ become equal (only zero winding sector
contributes) and are replaced by the momentum operator $p_y$ with
continuous spectrum.

The generalization of the superstring expressions in
\refs{\tserus,\rshet}\ is then as follows. Let $\hat N_R$ and
$\hat N_L$ denote number of states operators, which have the
standard form as in free superstring theory (in the NSR approach)
$ \hat N_{R,L}= N_{R,L} -a \ , \ \ \ a^{\rm (R)} =0 , \ a^{\rm
(NS) } =\ha $ (explicit expressions can be found in \refs{\rshet,
\rrt }). Let us introduce the angular momentum operators $\hat J_1
\equiv \hat J_{12}$ and $\hat J_2 \equiv \hat J_{34}$, which
generate rotations in the respective 2-planes (i.e. generate
shifts in $\vp_1 $ and $\vp_2$). They contain contributions of
both bosonic and fermionic oscillators for the respective
coordinates (of the same form as in flat space) and can be written
as follows: \eqn\eig{ \hat J_s=\hat J_{sL}+\hat J_{sR}\ ,\ \ \
\hat J_{sL} = l_{sL} + \ha + s_{sL} \ , \ \ \ \hat J_{sR} = -
l_{sR} - \ha + s_{sR} \ , \ \ \ \ \ \ s=1,2 \ , } where the
orbital momenta in each plane $l_{ L,R}=0,1,2,... $ are related to
the Landau quantum number $l$ and radial quantum number $k$ by
$l=l_L-l_R$ and $2k=l_L+l_R-|l|$, and $s_{ R,L}$ are the spin
components.
For the type II superstring theory, the spectrum for $R\to \infty
$ directly following from \rshet\ may be written as \eqn\wuno{
E^2-p_\a^2={2\over\a'} (\hat N_L +\hat N_R)+ p^2_y - 4
(p_y+E)(f_1\hat J_{1R}+f_2\hat J_{2R} ) \ . } Here $ \hat N_R-\hat
N_L=0 $ and the operators $N_{L,R}$ and $J_{sL,R}$ have the
standard flat-space form in terms of the bosonic and fermionic
creation/annihilation operators for the corresponding coordinates.
$p_\a$ are momenta in additional 4 spectator directions, and $E$
is conjugate to the time coordinate $ \te$.

As expected on the basis of the supersymmetry, the above spectrum
is tachyon free. This can be seen in a manifest way by writing
\wuno\ in the form \eqn\pwuno{ \big[ E+2(f_1\hat J_{1R}+f_2\hat
J_{2R} ) \big] ^2=p_\a^2+{2\over\a'} (\hat N_L +\hat N_R)+
\big[p_y - 2(f_1\hat J_{1R}+f_2\hat J_{2R} )\big]^2 \ , } which
shows that there is no state for which $E$ has an imaginary
component (note that $\hat N_L, \hat N_R=0,1,2,...$).

In the present context, it is natural to use the light-cone energy
$H=-p_u$, conjugate to the coordinate $u$ (which is actually the
original time of $AdS_3$, see \jeep). We have the relations
\eqn\dew{ p_u=\ha (p_y- E)\ ,\ \ \ \ p_v=\ha (p_y+E)\ ,\ \ \ p^u=
2 p_v\ ,\ \ \ p^v=2 p_u\ . } In terms of $p_u, p_v$ the \lc energy
spectrum takes the form \eqn\lic{ H =-p_u= { p_\a^2\ov 4 p_v} +
{1\over 2\a' p_v } (\hat N_L +\hat N_R) - 2 (f_1\hat
J_{1R}+f_2\hat J_{2R} ) \ , } i.e. \eqn\klic{ H = { p_\a^2\ov 4
p_v} + {1\over 2\a' p_v } \H \ ,\ \ \ \ \ \ \ \H= \hat N_L +\hat
N_R - 2 (\mm_1\hat J_{1R}+\mm_2\hat J_{2R} ) \ , \ \ \ \ \
\mm_{1,2} \equiv 2\a' p_v f_{1,2} \ . } It has the same form of
the Landau-type spectrum of a particle or open string in a
constant magnetic field (see, e.g., \tet\ for a review).

The partition function vanishes by virtue of supersymmetry of the
\pw background. \foot{In the bosonic theory, in the limit $R\to
\infty $ partition function becomes the same as the flat space
partition function.}

\medskip

The equivalent spectrum can be found also starting with \zs, i.e.
the same model written in the rotated coordinate system \hhoo. It
is convenient to introduce the complex coordinates $z_1= x_1 + i
x_2 =r_1 e^{i \vp'_1}$ and $z_2= x_3 + i x_4 =r_2 e^{i \vp'_2}$.
Then \zs\ becomes
$$ L = \del_+u \bd v + \sum_{s=1,2}
\big( - f^2_s z_s z_s^* \del_+u \bd u + \del_+z_s \bd z_s^* $$
\eqn\ccstro{ +\ \ha if_s \big[ \del_+u (z_s \bd z_s^*- z_s^* \bd
z_s)- \bd u (z_s \del_+z_s^* - z_s^* \del_+z_s) \big] \big) . }
The solution to the equations of motion can be easily found, e.g.,
by using the solutions of the model \fsto\ (explicitly given in
\refs{\tserus,\rshet}) and performing the coordinate
transformation \hhoo . In spite of the apparent mass term for the
bosonic fields $z_1,z_2$, the theory is again solved in terms of
massless free oscillators. Since the equation of motion for $u$ is
$ \del_+ \del_- u=0 $ we fix the light-one gauge $u= 2 \a' p^u
\tau = 4 \a' p_v \tau$. Then the equations for $z_1, z_2$ become
\eqn\smi{ \del_+\del_- z_s + \mm_s^2 z_s + i \mm_s ( \del_-z_s -
\del_+z_s)=0 \ , \ \ \ \ \ \mm_s \equiv 2 \a' p_v f_s \ , \ \ \ \
s=1,2 \ , } and thus are solved by \eqn\arsol{ z_s=e^{-2i\mm_s \s
}Z_s\ ,\ \ \ \ \ \ \ \ Z_s=Z_{s+}(\s_+) + Z_{s-}(\s_- )\ . }
Because $z_{1,2}(\s+\pi)=z_{1,2}(\s )$, the free fields $Z_{1,2}$
obey the twisted boundary conditions \eqn\bonn{ Z_1(\s+\pi
)=e^{2i\mm_1 \pi }Z_1(\s )\ ,\ \ \ \ Z_2(\s+\pi )=e^{2i\mm_2 \pi
}Z_2(\s )\ . } The mass spectrum is then found in the canonical
way, by computing the Virasoro operators and imposing the usual
Virasoro conditions for physical states (which are explicitly
solved in the \lc gauge).

A shortcut to the explicit form of the spectrum is to relate the
energy operator in the coordinate system corresponding to \cstro\
to the energy operator \wuno\ corresponding to \fsto. {}From the
coordinate transformation \hhoo\ (with $u'=u ,\ v'=v$), we learn
that $\hat J_s'=-i{\del\over \del\varphi_s'}=-i{\del\over \del
\varphi_s}=\hat J_s$ and \eqn\eeee{ E'=i{\del\over \del
t'}=i{\del\over \del t}-i f_1 {\del\over \del \varphi_1} -i f_2
{\del\over \del \varphi_2}=E+f_1 \hat J_1 +f_2\hat J_2 \ , }
\eqn\yyyp{ p'_y=-i{\del\over \del y'}=-i{\del\over \del y}+i f_1
{\del\over \del \varphi_1} +i f_2 {\del\over \del \varphi_2}=p_y
-f_1 \hat J_1 -f_2\hat J_2 \ . } Hence $p_u'=p_u-f_1 \hat J_1
-f_2\hat J_2 $, while $p_v$ does not change, $p_v'=p_v=\ha (p'_y +
E')= \ha (p_y + E)$. Thus the spectrum is given by \wuno\
expressed in terms of $E'$, $p'_y$: \eqn\nwuno{
{E'}^2-p_\a^2={2\over\a'} (\hat N_L +\hat N_R)+ {p'_y}^2-2
(p_y'+E')\big[f_1 (\hat J_{1R}-\hat J_{1L}) + f_2 (\hat
J_{2R}-\hat J_{2L})\big] \ , } supplemented by $ \hat N_R-\hat
N_L=0$. Equivalently, the analog of \lic\ is \eqn\nlic{ H\equiv
-p'_u= { p_\a^2\ov 4p_v} + {1\over 2\a' p_v } (\hat N_L +\hat N_R)
- f_1(\hat J_{1R}-\hat J_{1L})-f_2(\hat J_{2R}-\hat J_{2L}) \ , }
i.e. (cf. \klic) \eqn\nklic{ H= { p_\a^2\ov 4p_v} + {1\over 2\a'
p_v } \H \ ,\ \ \ \ \ \H= \hat N_L +\hat N_R - \mm_1(\hat
J_{1R}-\hat J_{1L})-\mm_2(\hat J_{2R}-\hat J_{2L}) \ . }
For $f_1=f_2=f$ this becomes simply \eqn\nnlic{ H= { p_\a^2\ov
4p_v} + {1\over 2\a' p_v } (\hat N_L +\hat N_R) - f(\hat
J_{R}-\hat J_{L}) \ , }
$$
\hat J_{R}\equiv \hat J_{1R}+\hat J_{2R} \ ,\ \ \ \ \ \ \ \hat
J_{L}\equiv \hat J_{1L}+\hat J_{2L} \ ,
$$
i.e. is related to $H$ in \lic\ (with $f_1=f_2$) by the shift
$H \to H + f (\hat J_{R}+ \hat J_{L})$. This  spectrum is
equivalent to the spectrum given in appendix C of  \mald\ (which
appeared while this paper was in preparation) with a proper
identification of modes.

Note that in spite of the presence of an apparent bosonic mass
term in \ccstro\ the spectrum has again the same ``magnetic'' form
as in \lic.

String theory in these magnetic backgrounds is periodic under
$\mm_1\to \mm_1+2n_1,\ \mm_2\to \mm_2+2n_2 $, with integer
$n_1,n_2$ (the periodicity in the case $\mm_2=0$ is discussed in
\refs{\tserus, \rshet }). Under such shifts, the spectrum remains
the same after  relabelling the mode operators. This is clear from
\nklic : the angular momentum components $\hat J_{1R}$, etc.,
appearing in this expression  have integer and half-integer
eigenvalues.  Shifts of  $\mm_1 ,\mm_2 $  by even integers can be
absorbed into integer shifts of $\hat N_R, \ \hat N_L$ which in
turn  can be  produced by a relabelling of mode operators (that
shifts the vacuum energy).

The case of  $\mm_1=\mm_2=\mm $ is special:  here the theory
becomes periodic under   $\mm \to \mm+ n$, with   $n $ being any
integer number.  This can be understood  by noting that the
eigenvalues of $\hat J_R$ are always  integer, so that a shift
$\mm \to \mm+ n$ can be absorbed into an integer shift of $\hat
N_R,  \hat N_L$. This  periodicity  follows also directly from the
form of the string action.
 Consider in particular the  ``left'' term
 in \oup, i.e.
  $\sim  S_L [ \del_0 - \del_1
+ \mm ( \g _{12} + \g _{34}) ] S_L $. Since $\g _{12} + \g _{34} =
2\g_{12}  P, \
 P \equiv \ha ( I - \g_{1234}), $ by splitting $S_L$ into two parts using
the projector $P$
(see also sect. 5.2), we
conclude that the  $\sigma $-dependence of the solution for the
$PS_L=S_L$  part is given by
 $\exp (2 \mm \sigma  \g^{12} ) $.
The  boundary condition for the  GS fermions  is the  periodicity
under
 $\sigma \to \sigma + \pi $;   this requirement  is
   invariant
under $\mm \to \mm +1$, determining   the periodicity of the
spectrum.
 In particular,
the string theory with $\mm_1=\mm_2=1$ is equivalent to the
theory with $\mm_1=\mm_2=0 $ (i.e.  to flat space). Without  loss
of generality, the values of $\mm $ can thus be restricted to the
interval $0\leq \mm <1 $.

%%%%%%%%%%%%%%%%%%%%%%%%%%%%%%
\newsec{Superstring model for \pw R-R 3-form background}
%%%%%%%%%%%%%%%%%%%%%%%%%%%%%

%%%%%%%%%%%%%%%%%%%%%%%%%
\subsec{ Lagrangian }
%%%%%%%%%%%%%%%%%%%%%%%%%

The S-dual to the above NS-NS background has the same plane wave
metric corresponding to the limit of the \ads now supported by the
R-R $F_3$ flux of the same null structure as $H_3$ (cf. \viv)
\eqn\zuup{ ds^2 = dudv - f^2 x_i^2 du^2 + dx_i^2 \ , } \eqn\suup{
F_{uij}= -\F_{ij} \ , \ \ \ \ \ \ \ \ \ \ \ \F_{12} = \F_{34} =2 f
\ . } In general, it is possible to argue that given the above \pw
metric supported by the R-R background $F_{u i_1...i_p} \sim
\F_{i_1...i_p}=$const, the only non-zero fermionic contribution to
the type IIB superstring action in the $\kappa$-symmetry \lc gauge
$\Gamma^u \theta^I=0$ is quadratic in fermions and comes from the
direct generalization of the fermion ``kinetic'' term in the
flat-space GS action \eqn\bass{i (\eta^{ab} \delta_{IJ} -
\epsilon^{ab} \rho_{3IJ}) \del_a x^{ m} \bar \theta^I \Gamma_{ m}
( \hat D_b)^{JK} \theta^K \ , } where $\theta^I$ are MW spinors
($I=1,2$) and $\rho_3=$diag(1,-1). The derivative $ \hat D_b$ is
the generalized covariant derivative that appears in the Killing
spinor equation (or gravitino transformation law) in type IIB
supergravity \schwarz. Acting on the real MW spinors $\theta^I$ it
has the form
%(we suppress the indices $I,J=1,2$)
\foot{We set to zero all background fields except $H_3$, $F_3$ and
$F_5$.} \eqn \derr{ \hat D_a = \del_a + { 1 \ov 4} \del_a x^k
\bigg[ (\om_{ m n k} - \ha H_{ m nk} \rho_3) \G^{ m n} - ( { 1 \ov
3!} F_{mnl} \G^{mnl} \rho_1 + { 1 \ov 2\cdot 5!} F_{mnlpq}
\G^{mnlpq}\rho_0 ) \G_k \bigg] } Here $ \rho_1 =\pmatrix{ 0 & 1
\cr 1 & 0}$ and $\rho_0 = \pmatrix{ 0 & 1 \cr -1 & 0} $ act in the
$I,J$ space.

The non-zero contribution to the action comes only from this
quadratic covariant derivative term where both $\del_a x^m $
factors in \bass\ are replaced by $\del_a u$, i.e. (after fixing
the bosonic \lc gauge $u= 2 \a' p^u \tau$) by $2 \a' p^u
\delta^m_u \delta_a^0$. In the flat space \lc gauge GS action the
spinors $\theta^1\equiv \tt_R$ and $\theta^2\equiv \tt_L$ become
the left and right moving 2-d fermions. The same happens in the
\pw background, and the surviving fermionic interaction terms are
proportional to $$ (\tt_L \G^{v} \G^{ij} \tt_L - \tt_R \G^{v}
\G^{ij} \tt_R) \F_{ij}$$ in the $H_3$ case ($\rho_3$ is diagonal),
and to $$\tt_L \G^{v} \G^{i_1...j_s} \tt_R \F_{i_1...j_s }$$ in
the $F_3$ and $F_5$ cases ($\rho_1$ and $\rho_0$ are
off-diagonal). In the $F_5$ case we reproduce the result \rrm\ of
the direct derivation of the GS action in the maximally
supersymmetric \pw background of \blau\ (see also \mt).

As expected, the fermionic interaction term has chiral (in 2-d
sense) structure in the NS-NS case, but the non-chiral,
``mass-term'' structure in the R-R case. This leads to important
differences in the properties (in particular, the spectrum) of the
NS-NS and the R-R models.

Explicitly, in the \lc gauge\foot{In what follows we shall use the
16-component spinors and $16\times 16$ $\gamma$-matrices instead
of 32-component spinors and $32\times 32$ $\Gamma$-matrices used
above. Note that $\G^u= \G^9 - \G^0 , \ \G^v= \G^9 + \G^0, \ \G^u
\G_u = 1$. } \eqn\gaul{ u = 2 \a' p^u \tau \ , \ \ \ \ \ \ \ \ \
\gamma^u \theta^I=0 \ , } where $\theta^I=(\tt_R,\tt_L)$ are
16-component MW spinors, we finish with\foot{ Here $\a=5,...,8$.
We assume that all $\g$-matrices are real and symmetric, so that
$M$ and $\g^v M$ are antisymmetric.} \eqn\lboy{L=L_B+ L_F \ , \ \
\ \ \ \ \ \ \ L_B= \del_+ u\del_- v - \mm^2 x_i^2 + \del_+ x_i
\del_- x_i + \del_+ x_\a \del_- x_\a \ , \ \ \ \ } \eqn\sett{ L_F=
i \theta_R{\gamma}^{v}\partial_+ \theta_R + i
\theta_L{\gamma}^{v}\partial_- \theta_L - 2 i \mm \theta_L
{\gamma}^{v} M \theta_R \ , } \eqn\demm{ \mm \equiv \a' p^u f = 2
\a' p_v f \ , \ } \eqn\saax{ M \equiv -{ 1 \ov 8 f} \F_{ij} \g^{ij
} = -\ha ( \g^{12} + \g^{34}) \ . } We have absorbed one power of
$p^u$ in \sett\ by the rescaling $\theta\to \sqrt{ p^u} \theta$.
While the $x_i$-part of the bosonic action has $SO(4)$ symmetry,
the corresponding symmetry of the fermionic action is only $SO(2)
\times SO(2)$, i.e. is the same as the symmetry of the $F_3$
background.

The resulting action \sett\ is essentially the same as found in
\rrm\ in the 5-form background case. As in the action of \rrm\ the
fermionic interaction plays the role of the 2-d fermion mass term,
mixing the left and right modes. Note, however, that here
\eqn\mmm{ M^2 = -\ha ( 1- \g^{1234}) } is (minus) a projector, in
contrast to the 5-form case \rrm\ where the corresponding mass
operator had its square proportional to 1. Thus only half of the
fermions are getting mass. The 4 massive + 4 massless bosons are
supplemented then by the 4 massive (2-d Majorana) fermions and 4
massless fermions. As in the flat GS action and the action of
\rrm\ here the \lc gauge GS action happens to have an effective
2-d supersymmetry.

The amount of the space-time supersymmetry here is the same as in
the ``S-dual'' NS-NS case. The \ads\ background with the R-R
3-form flux preserves 1/2 of the 32 supersymmetries, and its plane
wave limit also has 16 supersymmetries. Note that the restriction
$f_1=f_2$ is not necessary for the 1/2 of maximal supersymmetry of
the R-R plane wave model, but such model is a limit of the
corresponding \ads model where the equality of the radii of the
two 3-dimensional factors is crucial for the supersymmetry (and
also for the dilaton to be constant). The more general R-R \pw
model with field strengths $F_{u12} =-2f_1$, $F_{u34}=-2f_2$ can
be solved in a similar way as its $f_1=f_2$ case considered below.

%%%%%%%%%%%%%%%%%%%%%%%%%%%%%%%%
\subsec{Solution of the string model}

%%%%%%%%%%%%%%%%%%%%%%%%%%%%%%%%
\def\bmu{ {\Lambda } }
\def\bnu{ {\Sigma} }

The solution and quantization of this model is similar to the one
described in the R-R 5-form background case in \rf{\rrm,\mt}.

The equations of motion for $x_i$ and the fermions following from
\lboy,\sett\ are \eqn\xxxq{ \del_+ \del_- x_i + \mm ^2 x_i=0\ , \
\ \ \ \ \ \ \del_+\del_- x_\a=0\ , } \eqn\ththq{\del_+\theta_R -
\mm M \theta_L=0 \ ,\ \ \ \ \ \ \ \ \ \ \del_-\theta_L- \mm M
\theta_R=0 \ . } As in flat space case, let us solve the \lc gauge
condition on fermions \gaul\ explicitly, expressing $\theta_{L,R}$
in terms of 8+8 independent real fermions $S_{L,R}$ (spinors of
$SO(8)$). One can choose the following representation for the
$16\times 16$ gamma matrices:
$$
\gamma^u \to I \otimes \s^+\ ,\ \ \ \gamma^v\to I \otimes \s^- \ ,
\ \ \ \ \ \ \ \gamma^i\to \g^i \otimes \s^3 \ ,
$$
where the new $\g^i$'s are $8\times 8$ gamma matrices of $SO(8)$
group. $\s$'s are Pauli matrices and $I$ is the $8\times 8$
identity matrix. In this representation one is left with the
upper-component spinor $S$, and the 16-component and 8-component
gamma matrices with transverse indices are directly related, i.e.
$\gamma^{12}\to \g^{12} , \ \gamma^{34}\to \g^{34}.$ Then \ththq\
takes the same form but now in terms of unconstrained 8-component
spinors. In particular, we learn that \eqn\err{ \del_+\del_-
S_{L,R} - \mm ^2 M^2 S_{L,R}=0\ , } where $M$ now is $8 \times 8$
matrix of the same structure as in \saax.

Let us further split the fermions in the 8$\to$4+4 way $S_L\to
(S_L,\hat S_L)$, $S_R\to (S_R,\hat S_R)$, so that \eqn\wer{
\gamma^{1234} \pmatrix{S_{L,R}\cr \hat S_{L,R}}= \pmatrix{
-S_{L,R} \cr \hat S_{L,R}}\ . } One may solve the constraints in
\wer\ explicitly by reducing the spinors to independent
4-component ones. For example, diagonalizing $\g^{12}$ so that
\eqn\ded{ \gamma^{12} \pmatrix{S_{L,R}\cr \hat S_{L,R}}=
-\pmatrix{ \bmu S_{L,R} \cr \bnu \hat S_{L,R}}\ , \ \ \ \ \ \ \ \
\ \bmu^2=\bnu^2=-I \ , }
 where $\bmu$ and $\bnu$ are $4 \times 4$
traceless antisymmetric matrices with eigenvalues $\pm i$.
we find 
\eqn\otra{ M \pmatrix{S_{L,R}\cr \hat S_{L,R}}= \pmatrix{ \Lambda
S_{L,R} \cr 0 }\ , } and eqs. \ththq\ take the form \eqn\ehhh{
\del_+ S_{R} - \mm \Lambda S_{L}=0\ ,\ \ \ \ \ \ \ \ \del_- S_{L}-
\mm \Lambda S_{R}=0\ , } \eqn\eggg{ \del_+ \hat S_{R} =0\ ,\ \ \ \
\ \ \ \ \ \ \del_- \hat S_{L} =0\ . \ } We are left with 4+4
independent degrees of freedom $(S^a_L,\hat S^A_L)$, $(S^a_R,\hat
S^A_R)$, where $a=1,2,3,4$ and $A=1,2,3,4$. In what follows the
explicit structure of the matrix $\bmu$ (i.e. the upper $4\times
4$ block of $\g^{12}$) will not be important -- only it square
$(=-I)$ will appear in the resulting Hamiltonian.
%It is also useful to note that in the present
%Majorana representation one has $\Lambda^{\bf T}=-\Lambda $,
%$\Sigma^{\bf T}=-\Sigma $.

Note that $S^a_{L,R}$ do not transform as an $SO(4)$ spinor under
rotations in $(1,2,3,4)$ directions: the presence of $\g^{12}$ in
its definition \ded\ breaks the $SO(4)$ Lorentz symmetry down to
$SO(2) \times SO(2)$, with the latter being the true global
symmetry of the R-R 3-form background \suup, i.e. of the fermionic
sector of the string action \sett, and thus of the whole string
theory (see \mt\ for a related discussion in the 5-form background
case).

Thus we finish with four massive 2-d bosons $x_i$ and four massive
fermions $S^a_{L,R}$ of the same mass $\mm$, as well as with four
massless free bosons $x_\a$ and four massless fermions $\hat
S^A_{L,R}$. \foot{In what follows we will omit the hat over $S^A$,
differentiating between $S$ and $\hat S$ by type of their
indices.}

The Fourier expansion for generic closed-string classical
solutions $x^i$ can be written as
$$ x^i(\s,\tau ) = {i
l\over 2} \bigg[ {1 \ov \sqrt{w_0}} \big( a_0^i e^{-2iw_0\tau } -
a_0^{i\dagger} e^{2iw_0 \tau}\big) + \sum_{n=1}^\infty {1\over
\sqrt{w_n}}\big[ e^{-2i w_n\tau }\big( a_n^i e^{2in\s } + \tilde
a_n^i e^{-2in\s } \big) $$ \eqn\foxxx{- \ e^{2i w_n\tau }\big(
a_n^{i\dagger } e^{-2in\s } + \tilde a_n^{i\dagger} e^{2in\s }
\big)\big] \bigg] \ , } where \eqn\wsa{ w_n=\sqrt{n^2+ \mm ^2 }\
,\ \ \ \ \ w_0 = \mm \ , \ \ \ \ \ \ \ l\equiv \sqrt{2\a' }\ . }
Note that $\mm $ and $w_n$ are dimensionless; we have introduced
the string scale $l$ to make the Fourier mode coefficients
dimensionless. The canonical commutation relation $[\dot x^i(\s
),x^j(\s ')]=-2\pi\a 'i \delta (\s-\s ')\delta^{ij}$ then implies
as usual \eqn\rejj{ [a_n^i,a_m^{j\dagger }]= \delta_{nm}
\delta^{ij}\ ,\ \ \ \ \ [\tilde a_n^i,\tilde a_n^{j\dagger
}]=\delta_{nm} \delta^{ij}\ . } In addition, we have massless
boson modes $a^\a_n,\tilde a^\a_n$ appearing in the expansion of
$x^\a$. For the fermions, the solutions are
$$ S_R^a(\s ,\tau )= l \bigg[
c_0 \big( S_0^a e^{-2iw_0\tau } + S_0^{a\dagger} e^{2iw_0
\tau}\big) + \sum_{n=1}^\infty c_n \big[ e^{-2i w_n\tau }\big(
S_n^a e^{2in\s } + \tilde S_{n}^{'a} e^{-2in\s } \big)
$$ \eqn\sssr{
+\ e^{2i w_n\tau }\big( S_n^{a\dagger } e^{-2in\s } + \tilde
S_n^{'a\dagger} e^{2in\s } \big)\big]\bigg] \ , }
$$ S_L^a(\s ,\tau )= l\bigg[ c_0 \big( \tilde S_0^a
e^{-2iw_0\tau } + \tilde S_0^{a\dagger} e^{2iw_0 \tau}\big) +
\sum_{n=1}^\infty c_n \big[ e^{-2i w_n\tau }\big( \tilde S_n^a
e^{-2in\s } + S_{n}^{'a} e^{2in\s } \big)
$$
\eqn\sssr{ +\ e^{2i w_n\tau }\big( \tilde S_n^{a\dagger } e^{2in\s
} + S_n^{'a\dagger} e^{-2in\s } \big)\big]\bigg] \ , } where
$c_0,c_n$ are the normalization constants to be fixed below. The
primed fermion modes $S^{'a}_n, \tilde S^{'a}_n $ are related,
according to to the first-order equation \ehhh, to $S^{a}_n,
\tilde S^{a}_n, $ by which gives \eqn\quy{ \tilde S_{n}^{'a} =
{i\over \mm }(w_n-n) (\bmu \tilde S_n)^a\ ,\ \ \ \ \tilde
S_{n}^{'a\dagger} = -{i\over \mm }(w_n-n) (\bmu \tilde
S^\dagger_n)^{a }\ ,\ } \eqn\quyy{ S_{n}^{'a} = {i\over \mm
}(w_n-n) (\bmu S_n)^a\ ,\ \ \ \ S_{n}^{'a\dagger} = - {i\over \mm
}(w_n-n) (\bmu S^\dagger_n)^{a }\ ,\ } and \eqn\zzuy{ S_0^a=i(\bmu
\tilde S_0)^a\ ,\ \ \ \ \ \ \ \ \ S_0^{a\dagger }=-i(\bmu \tilde
S_0^{\dagger })^a\ . } In the zero-field limit $\mm\to 0$ (i.e.
$f\to 0$) we recover the standard flat-space Fourier expansions.
Imposing the canonical commutation relations \eqn\ghuy{ \{
S_L^a(\s,\tau ), S_L^b(\s ',\tau )\} =2\pi\a '\delta^{ab}\delta(\s
-\s ')\ ,\ \ \ \ \{ S_R^a(\s,\tau ), S_R^b(\s ',\tau )\} =2\pi\a
'\delta^{ab}\delta(\s -\s ')\ , \ } and fixing the normalization
constants as \eqn\vvv{ c_0={1\over \sqrt{2}}\ ,\ \ \ \ \ \ \ \ \
c_n={\mm \over \sqrt{\mm ^2+(w_n-n)^2}}\ , } we get the following
anti-commutation relations \eqn\zcan { \{ S_m^a, S_n^{b\dagger}\}
= \delta^{ab} \delta_{mn}\ ,\ \ \ \ \{ \tilde S_m^a, \tilde
S_n^{b\dagger}\} = \delta^{ab} \delta_{mn} \ ,\ \ \ \ \ \ \ \{
S_0^a, S_0^{b\dagger}\} =\delta^{ab}\ . } The massless fermions
$S^A_{L,R}$ have the standard Fourier expansion in terms of the
corresponding fermion modes $S_n^A,\tilde S_n^A$ satisfying $\{
S^{A} _n, S_m^{B\dagger}\} =\delta^{AB} \delta_{nm}$.

We shall define the vacuum $|0\rangle $ as a state annihilated by
$a_n^i, \tilde a_n^i$ and $S^a_n , \tilde S^a_n $, with
$n=1,...,\infty $, as well as by $a_0^i$ and $S^a_0$. This is a
natural definition, since, in particular, it ensures the
regularity at $-\infty$ of the euclidean time axis: the physical
states created by acting with vertex operators on $|0\rangle $ at
$\tau \to i\infty $ must be regular, and therefore the
coefficients of $e^{-2iw_n\tau }$ in the Fourier expansion must
annihilate the vacuum.\foot{The same definition of the vacuum was
assumed in \mald. As discussed in \mt, the definition of the
vacuum state in the zero-mode sector is not, in fact, unique, with
different definitions corresponding to different orderings of the
states in the zero-mode supermultiplet.}
The representation of the Clifford algebra satisfied by the
fermionic zero modes is constructed by acting by $S_0^\dagger$ on
the Clifford vacuum state.

The light-cone Hamiltonian can be written as
$$ H=-p_u={1\over 16\pi \a'
p_v}\int_0^\pi d\s \bigg(\del_0 x_i\del_0 x_i+\del_1 x_i\del_1
x_i+ 4 \mm ^2 x_i^2 +\del_0 x_\I\del_0 x_\I+\del_1 x_\I \del_1
x_\I
$$ \eqn\hhee{
-\ i S_R^a\del_1 S_R^a + i S_L^a \del_1 S_L^a + 4i \mm S_L^a
\bmu_{ab} S^b_R -i S_R^A\del_1 S_R^A +i S_L^A \del_1 S_L^A \bigg)
\ . } By using the equations of motion for the fermions, we can
put it in the form $$ H= {1\over 16\pi \a' p_v} \int_0^\pi d\s
\bigg(\del_0 x_i\del_0 x_i+\del_1 x_i\del_1 x_i+ 4 \mm^2 x_i^2
+\del_0 x_\I\del_0 x_\I+\del_1 x_\I \del_1 x_\I
$$\eqn\heep{
+\ i S_R^a\del_0 S_R^a + i S_L^a \del_0 S_L^a + i S_R^A\del_0
S_R^A +i S_L^A \del_0 S_L^A \bigg) \ . } This gives \eqn\hhar{
H=-p_u ={p_\a^2\over 4p_v}+{1\over 2\a'p_v}\H \ , \ \ \ \ \ \ \ \H
=\H_0+ \H_R+ \H_L + N^0_R+ N^0_L \ , } where we have defined
($w_n=\sqrt{\mm^2 + n^2}, \ n=0,1,2,...$) \eqn\nnj{ \H_0=w_0
\big(a^{i\dagger}_{0} a_{0}^i + S^{a\dagger}_0 S^a_0 \big) \ , }
\eqn\hyt{ \H_R= \sum_{n=1}^\infty w_n \big(a^{i\dagger}_{n}
a_{n}^i + S^{a\dagger}_n S^a_n \big) \ ,\ \ \ \
\H_L=\sum_{n=1}^\infty w_n \big( \tilde a^{i\dagger}_{n} \tilde
a_{n}^i + \tilde S^{a\dagger}_n \tilde S^a_n \big) \ , }
\eqn\ffdd{ N_R=\sum_{n=1}^\infty n \big( a^{i\dagger}_{n} a^i_{n}
+ S^{a\dagger}_n S^a_n \big) \ , \ \ \ \ N_L=\sum_{n=1}^\infty n
\big( \tilde a^{i\dagger}_{n} \tilde a^i_{n} + \tilde
S^{a\dagger}_n \tilde S^a_n \big)\ , } \eqn\nnjd{
N_R^0=\sum_{n=1}^\infty n \big( a^{\I\dagger}_{n} a_{n}^\I +
S^{A\dagger}_n S^A_n \big) \ ,\ \ \ \ N_L^0=\sum_{n=1}^\infty n
\big( \tilde a^{\I\dagger}_{n} \tilde a^\I_{n} + \tilde
S^{A\dagger}_n \tilde S^A_n \big) \ . } The physical states must
satisfy the constraint (expressing invariance under translations
in $\s$) \eqn\consu{ N_R+ N_R^0= N_L+ N_L^0 \ . } Note that with
the definition of the vacuum state we have assumed, the normal
ordering constants have cancelled between the fermionic and
bosonic contributions, both in the string-oscillator and the
zero-mode sectors. The same Hamiltonian was given in \mald.

The string states are then constructed by acting by the creation
operators on the vacuum. Apart from the translational contribution
${p_\a^2\over 4p_v}$ in 4 directions $x_\I$ this results in a
discrete spectrum of states.

%%%%%%%%%%%%%%%%%%%%
\subsec{Equivalence of the supergravity parts of the spectra of
the NS-NS and R-R models}
%%%%%%%%%%%%%%%%%%%%%%%%%%%%%%

The NS-NS and R-R \pw models correspond to the two S-dual 3-form
backgrounds. The S-duality is an exact global symmetry of the type
IIB supergravity \schwarz, and thus the spectra of fluctuations of
the supergravity fields near the two S-dual backgrounds must be in
one-to-one correspondence.\foot{The same applies of course to the
case of the $AdS_3 \times S^3$ NS-NS and R-R backgrounds.} That
implies that the zero-mode (i.e. the $\a'\to 0$) parts of the
spectra \nnlic\ and \hhar\ of the two string models we have
discussed above must be equivalent.

That the two string models are indeed equivalent in the
point-particle ($\a'\to 0$) limit follows directly from the form
of the corresponding \lc string actions. The bosonic parts of the
two actions (cf. \zro\ and \zuup,\lboy) have the same target space
metric part, while the extra $B_{mn}$-coupling term in the bosonic
action does not contribute in the point-particle limit. The
fermionic actions \oup\ and \sett\ are related by a rotation of
the fermions: the $H_{mnk}$ and $F_{mnk}$ terms in the covariant
derivative \derr\ are related by the transformation of $\theta^I$
(or $S_L,S_R$) that rotates the matrix $\rho_3 = \pmatrix{ 1 & 0
\cr 0 & -1} $ into $\rho_1 =\pmatrix{ 0 & 1 \cr 1 & 0}$; the
difference between the kinetic terms of the ``left'' and ``right''
fermions disappears in the particle limit ($\del_+ \to \del_-\to
\ha \del_0$).

It is easy to see then the two zero-mode partition functions $Z_0
= $Tr$ e^{-\b \H_0}$ coincide, and are equal to $Z_0 = 16
\big({1+e^{-\bb\mm }\over 1-e^{-\bb\mm } }\big)^4 $. The bosonic
contributions are obviously the same. The fermionic contribution
follows easily from \nnj. To reproduce it in the NS-NS case one
may use the GS formulation where the action is given by \oup\ and
the $\H_0$ part of \nnlic\ is $ \H_0 = \mm (\hat J_L -\hat J_R)$
where in GS representation the zero-mode parts of the angular
momentum operators $J_{L,R}$ are given by the $\sim S_0 (\g^{12} +
\g^{34}) S_0$ combinations. Doing the same splitting into
$(S^a,S^A)$ as above (cf. \ded) and rotating the fermions one
arrives at the same result for $Z_0$ as in the R-R case.

The equality between the supergravity parts of the respective
partition functions shows that in the supergravity sector of the
the NS-NS and R-R models  there is an equal number of states for
each value of the \lc energy. It is instructive to see directly
from the explicit form of the expressions for the NS-NS \nnlic\
and the R-R spectra \nnj\ how different states are mapped into
each other. {}From \nnlic\ we get (setting $\hat N_L=\hat N_R=0$
and omitting the free translational part) \eqn\nsnsn{ \H_{0}^{\rm
NS-NS} =\mm (\hat J_L-\hat J_R) \ .} Using eq.~\eig , we find that
the eigenvalues are \eqn\qaap{ \E_{0}^{\rm NS-NS}= \mm (
l_{1L}+l_{1R}+l_{2L}+l_{2R} ) +k_{\rm NS-NS}\ , \ \ \ \ k_{\rm
NS-NS}\equiv \mm (2+ s_{1L}+s_{2L} -s_{1R}-s_{2R}) , } where
$l_{L,R}=0,1,2,...$, and (for states with $\hat N_L=\hat N_R=0$)
$s_{L,R}$ can take values $-1,-{1\over 2},0,{1\over 2},1$. They do
not take these values {\it independently}, because of the
condition $\hat N_L=\hat N_R=0$. For example, in the NS-NS sector,
where $N_L=1,\ N_R=1$, their possible values satisfy the condition
$ |s_{1R}\pm s_{2R}| \leq 1\ ,\ \  |s_{1L}\pm s_{2L}| \leq 1\ . $
On the other hand, the R-R expression \nnj\ is \eqn\rrrr{
\H_{0}^{\rm R-R}=\mm (a_0^{i\dagger }a_0^i +S_0^{a\dagger }S_0^a )
\ . } The occupation number of the modes $a_0^i$, $i=1,2,3,4$, can
be any value from 0 to $\infty $, so we may call them
$l_{1L},l_{1R},l_{2L},l_{2R}$. Let us call $s_0^a $ the occupation
number of the fermion modes $S_0^a$, where $s_0^a=0,1$. Thus the
energy eigenvalues are \eqn\kkrrr{ \E_{0}^{\rm R-R}= \mm
(l_{1L}+l_{1R}+l_{2L}+l_{2R})+k_{\rm R-R}\ , \ \ \ \ \ \ \ k_{\rm
R-R} \equiv \mm (s_0^1+s_0^2 +s_0^3+s_0^4) \ . } It is easy to see
that the contributions $k_{\rm R-R}$, $k_{\rm NS-NS}$ indeed take
the same values for the different states of the supergravity part
of the spectrum. For $k_{\rm R-R}$, the possible values are $0,
\mm ,2\mm ,3\mm ,4\mm $. For $k_{\rm NS-NS}$, the possible values
are indeed the same. The minimal $k_{\rm NS-NS}$ is for the states
with $s_{1R}=1, s_{2R}=0 , s_{1L}=-1, s_{2L}=0 $, and other three
similar states obtained by  exchanging the indices
$1\leftrightarrow 2$. The maximal value of $k_{\rm NS-NS}$  is
found for similar states with reverse sign  of  the spin. $k_{\rm
NS-NS}$ takes only integer values, because $s_{1R}+s_{2R}$ and
$s_{1L}+s_{2L}$ are always integer. Thus one sees that the
possible values are $k_{\rm NS-NS}=0, \mm ,2\mm ,3\mm ,4\mm $.  To
find multiplicities, one should also take into account the extra
free oscillators. In the NS case, the multiplicity counting is
more complicated, but the above proof that $Z_0^{\rm R-R}=Z_0^{\rm
NS-NS}$ implies that multiplicities are the same. Thus we conclude
that the supergravity parts  of the NS-NS and R-R spectra match
exactly, as required by S-duality.

%%%%%%%%%%%%%%%%%%
\newsec{Asymptotic density of states in \pw string models }
%%%%%%%%%%%%%%%%%%

The spectra of the models discussed above have several interesting
features. Here we will discuss the corresponding density of states
at given \lc energy. We shall evaluate its asymptotic behavior at
large energy.
This may be viewed also as a first step towards a study of
thermodynamics of the corresponding string ensembles.\foot{To set
up thermodynamics analysis may be subtle (in particular, the \pw
background does not admit a straightforward euclidean-time
continuation -- the \pw metric \zuup\ becomes complex).
Here we shall simply study
the number of string states with given \lc energy at fixed $p^u$.
}

We shall use the dimensionless \lc energy $\E$ measured in units
of $2\a' p_v$, i.e. the eigenvalue of $\H$ in \klic,\nklic,\hhar.
In general, the total number $d_{\E }$ of physical string states
with given \lc energy $\E$ can be found from
\eqn\dee{ Z =\sum_\E d_{\E }\ \nu^\E = \sum_\E d_{\E }\ e^{-\bb \E
} \ , } \eqn\deed{ d_\E ={1\over 2\pi i}\oint {d\nu\over \nu ^{\E
+1}}\ Z(\nu) \ ,\ \ \ \ \ \ \ \ \ \nu\equiv e^{-\bb } \ . } Here
$\bb$ is dimensionless ``inverse temperature'' (related to the
standard temperature by $\b \to { \b\ov 2 \a' p_v}$) and the
partition function $Z$ is defined by \foot{This $Z$ is the \lc
gauge (``transverse-mode'') partition function of a single string.
To obtain a thermal partition function for a gas of strings in
flat space \alos\ one needs to integrate over $p^+\equiv
p^u=2p_v$, i.e. to compute $Z_1 (\bb) = \int^\infty_0 dp_v e^{-\b
p_v} Z( { \bb \ov 2 \a' p_v}) $ (related to the trace of $e^{-\b
p_0}, \ p_0= p_v - p_u= p_v + H$) and then to further
``exponentiate'' it accounting for the statics of states, Z$(\bb)
= \exp [ \ha \sum_{r=1}^\infty {1 \ov r} ( 1 - (-1)^r) Z_1
(r\bb)]$. } \eqn\qual{ Z={\rm Tr}\left( e^{-\bb \H}\right) =
\int_0^{2\pi }d\lambda \ {\rm Tr'} \left( e^{-\bb \H} e^ {i\lambda
( \hat N_R- \hat N_L) }\right) \ . } The integral over $\lambda$
imposes the level matching constraint (the analog of \consu), i.e.
the trace Tr is the sum over the physical states, whereas Tr$'$ is
the sum over all states of the Fock space.\foot{ In the flat space
case ($f=0$) the constraint can be imposed at the end of the
calculation by first treating the left and right sectors
separately and then setting $\H_R=N_R$ equal to $\H_L=N_L$ in the
definition of $d_{\E_R,\E_L}$. Here we need to impose the level
matching constraint by integrating over the Lagrange multiplier
$\lambda $ since $\H_{R,L} $ are no longer equal to $N_{R,L}$ (cf.
\nnj,\ffdd).}

Since we shall be interested in the $\E \to \infty$ asymptotic
form of $d_\E$, it will be sufficient to evaluate $Z$ for $\bb \to
0$ (i.e. $\nu\to 1$). Having found $Z$, the integral in \deed\ can
be computed in a saddle point approximation.

%%%%%%%%%%%%%%%%%%%%
\subsec{Models with R-R background}
%%%%%%%%%%%%%%%%%%%%%%%%%%%%%%
We shall first consider the \pw models with the 3-form and 5-form
R-R backgrounds. In the case of the 3-form model with the spectrum
\hhar\ we obtain \eqn\zual{ Z = Z_0 Z_{str} \ ,\ \ \ \ \ \ Z_0 =
16 \big({1+e^{-\bb\mm }\over 1-e^{-\bb\mm } }\big)^4 \ , }
\eqn\sre{ Z_{str}= \int_0^{2\pi }d\lambda \ \prod_{n=1}^\infty
\bigg|{1+e^{-\bb w_n+i\lambda n }\over 1-e^{-\bb w_n+i\lambda n}
}\bigg|^8 \prod_{n=1}^\infty \bigg|{1+e^{-\bb n+i\lambda n }\over
1-e^{-\bb n+i\lambda n} }\bigg|^8 \ . }
In the flat space limit $f\to 0$, i.e. $w_n \to n$, \sre\ becomes
the standard superstring partition function which can be expressed
in terms of $\theta$-functions. In computing the zero-mode
contribution $Z_0$ we have omitted the translational zero-mode
term $p^2_\I\ov 4p_v$ in \hhar. The factor 16 in $Z_0$ accounts
for the degeneracy in the zero mode part of the ``free'' sector
($S^A,x^\I)$. The zero-mode contribution gives $Z_0({\bb \to 0} )
\to 16 ({2 \ov \b\mm})^4$.

To evaluate the string-mode contribution $Z_{str}$ in the ``large
temperature'' limit $\bb \to 0$ we shall first ignore the
$\l$-dependence (i.e. ignore the level matching constraint) and
discuss its effect later. Let us formally define (cf.\sre)
\eqn\zrig{ Z_R=Z_L=\prod_{n=1}^\infty \bigg({1+e^{-\bb w_n}\over
1-e^{-\bb w_n} }\bigg)^4 \prod_{n=1}^\infty \bigg({1+e^{-\bb n
}\over 1-e^{-\bb n} }\bigg)^4 \ , } and consider the following
``bosonic'' factor in \zrig: $\prod_{n=1}^\infty (1-e^{-\bb w_n
})\equiv e^{K}$, where \eqn\groz{ K(\bb,\mm) \equiv
\sum_{n=1}^\infty \log (1-e^{-\bb w_n }) =- \sum_{n,k=1}^\infty
{1\over k} e^{-\bb k w_n} \ . } To extract the leading $\bb\to 0 $
behavior, we have to sum over $n$ and identify the terms which
diverge as $\bb\to 0 $. Let us expand $ w_n=\sqrt{n^2+\mm ^2}
\cong n + {\mm ^2\over 2n}+O({\mm ^4\ov n^3}). $ Then \eqn\yyy{
e^{-\bb k w_n}\cong e^{-\bb k n} e^{-{\bb k \mm^2\over 2n} }\cong
e^{-\bb k n} (1-{\bb k \mm ^2 \over 2n} +...) \ , } where dots
represent terms which become
%(after suming over $n$)
$\bb $-independent in the limit of small $\bb $. Using the
symmetry between the sums over $k$ and $n$ we get \eqn\zaz{ K
\cong - \sum_{n,k=1}^\infty e^{-\bb k n} ( {1\over k} -{\bb \mm ^2
\over 2n} +...) =
%- \sum_{k=1}^\infty \big[ {e^{-\bb k}\over
%k(1- e^{-\bb k})} +\ha \bb \mm ^2 \log (1-e^{-\bb k})\big]
-(1-\ha \bb \mm^2 )\sum_{k=1}^\infty {e^{-\bb k}\over k(1- e^{-\bb
k}) } + ... \ , } so that for $\bb \to 0$ \eqn\fnal{K \cong
-{\pi^2\over 6\bb } +{\pi^2 \mm^2\over 12} + O(\mm^ 4)+ O(\b) \ .
} We have used that $\sum_{k=1}^\infty {1\over k^2} ={\pi^2\over
6}$. Thus the leading term at small $\bb $ is independent of $\mm
$, while the subleading contribution does not depend on $\bb $.

In general, $K$ in \groz\ can be written as a series of Bessel
functions. The resulting expression has simpler form if the zero
mode contribution (see \zual ) is included in the sum. Using the
Poisson formula, we get \eqn\zez{ \sum_{n=-\infty }^\infty \log
(1-e^{-\bb w_n }) = \sum_{k=-\infty }^\infty \int dx \ e^{2\pi i x
k} \log(1-e^{-\beta w_x})\ , \ \ \ w_x=\sqrt{x^2+\mm ^2} \ . }
Expanding the logarithm and performing the integration we obtain
\eqn\bess{ \sum_{n=-\infty }^\infty \log (1-e^{-\bb w_n }) =-
2\sum_{k=-\infty }^\infty \sum_{p=1}^\infty {\mm \beta \over
\sqrt{\beta ^2 p^2 + 4\pi ^2 k^2}} {\rm K}_1(\mm \sqrt{\beta^2 p^2
+4\pi^2 k^2}) \ , } where ${\rm K}_1$ is the modified Bessel
function.

To find the explicit $\mm $ dependence of $K$ in the small $\bb$
limit let us note that
\def\mapa#1{\smash{\mathop{\longrightarrow }\limits_{#1} }}
\eqn\weret{ {\del K\over \del \mm } = \bb \mm \sum_{n=1}^\infty
{e^{-\bb w_n}\over w_n(1-e^{-\bb w_n}) } \ \ \ \mapa{\beta\to 0} \
\ \ \mm \sum_{n=0}^\infty {1\over w_n^2 }={ \pi \ov 2} \coth \pi
\mm - { 1 \ov 2 \mm } \ , } so that
$$
K(\bb , \mm )\cong K_0(\bb ) + \ha \log { \sinh \pi \mm \ov \pi
\mm} + ... \ .
$$
Expanding in powers of $\mm $, one finds
$$e^{K}\cong e^{K_0(\beta )} \ \bigg({\sinh \pi \mm \over \pi \mm }\bigg)^{1/2} \cong e^{K_0} (1 + {\pi^2
\mm^2\over 12 }+...) \ , $$ where the second $O(\mm^2 )$ term
corresponds to the correction derived in \fnal. The factor
$e^{K_0(\bb )}$ is the standard $\bb $ dependent factor of the
flat-space ($\mm =0$) theory, with the leading behavior
$e^{K_0}\cong \beta^{-1/2}\ e^{-{\pi^2\over 6 \bb }}$ (cf. \fnal
).

The analog of $K$ \groz\ corresponding to the fermionic
contribution in \zrig\ is $K_F(\bb ,\mm )=\sum_{n=1}^\infty
\log(1+e^{-\bb w_n})$. It can be analysed in a similar way as in
\zaz, \weret; we find that in this case there is no $\mm
$-dependent factor in the $\bb \to 0 $ limit. As a result, we find
\eqn\zzzjj{ Z_R \cong {\rm const.}\ \bb ^{6} e^{2\pi ^2/\bb } \
\left( {\pi \mm \over \sinh \pi \mm} \right)^2\ . }
It is easy to see that including the dependence on the Lagrange
multiplier $\lambda $ in \sre\ does not change the leading
dependence on $\bb $. Now instead of \yyy\ we have
$$
e^{-\bb k w_n+i\lambda kn}\cong e^{-\bb k n+i\lambda kn} e^{-{\bb
k \mm^2\over 2n} }\cong e^{-\bb k n+i\lambda kn} (1-{\bb k \mm ^2
\over 2n} +...) \ .
$$
The second $\mm^2$ term in the brackets is irrelevant for the
leading $\bb \to 0$ behavior: as above, it gives a finite
contribution.
Thus the density of states as a function of the energy is expected
to have the same leading large $\E$ behavior as in the flat-space
theory, \ \ $d_\E\sim e^{4\pi \sqrt{\E }}$. It would be
interesting to determine explicitly the $\mm$-dependence of the
subleading terms in $d_\E$.

The same analysis can be repeated for the string model of
\refs{\rrm,\mald,\mt} corresponding to the \pw background \blau\
with the R-R 5-form field of strength $f$. Here the \lc
Hamiltonian can be obtained from \hhar\ by replacing 4+4 by 8+8
massive bosons and fermions and discarding the 4+4 massless
species (so that now $i=1,...,8; \ a=1,..., 8$ and the are no
$N^0_{L,R}$ and $p_\I^2$ terms in \hhar,\consu). The corresponding
string spectrum was described in detail in \mt.
Explicitly, one finds \eqn\her{ H \equiv (2 \a' p_v)^{-1}
 \H \ ,\ \ \ \ \
\ \ \H = \H_0 + \H_R + \H_L \ , \ \ \ \ \ \ \H_0 = \mm(
a^{i\dagger}_{0} a_{0}^i + S^{a\dagger}_0 S^a_0 ) \ ,} \eqn\xde{
\H_R = \sum_{n=1}^\infty w_n \ \big(a^{i\dagger}_{n} a_{n}^i +
S^{a\dagger}_n S^a_n) \ ,\ \ \ \ \ \ \H_L = \sum_{n=1}^\infty w_n
\big(\tilde a^{i\dagger}_{n} \tilde a^i_{n} + \tilde
S^{a\dagger}_n \tilde S^a_n \big) \ ,} \eqn\fer{ w_n \equiv
\sqrt{n^2 + \mm^2} \ , \ \ \ \ \ \ \ \ \ \mm = 2 \a' p_v f \ . }
The $\mm=0$ limit is the flat string theory limit, while the $\a'
\to 0$ limit is the supergravity (zero-mode) limit where all
string oscillation modes become infinitely heavy and decouple.
This Hamiltonian has also another regular limit \mt: $\a' \to
\infty$ or $\mm \to \infty$, i.e. the tensionless string limit,
where $ w_n \to \mm$. In this limit there is the same mass scale
$\mm$ at all string levels including the zero-mode one: at all
levels there are states with the same mass as at the lowest
(supergravity) level so that one cannot naturally decouple the
string states from the supergravity states.

Here we find that $Z$ in \qual\ takes the form (cf. \zual,\sre)
\eqn\ual{ Z = Z_0 Z_{str} \ ,\ \ \ \ \ \ Z_0 = \big({1+e^{-\bb\mm
}\over 1-e^{-\bb\mm } }\big)^8 \ , \ \ \ \ \ \ Z_{str}=
\int_0^{2\pi }d\lambda \ \prod_{n=1}^\infty \bigg|{1+e^{-\bb
w_n+i\lambda n }\over 1-e^{-\bb w_n+i\lambda n} }\bigg|^{16} \ . }
In the small $\bb $ limit, we get the following estimate for the
``unconstrained'' right part of $Z_{str}$ (i.e. the analog of
\zrig) \eqn\zzzj{ Z_R\cong {\rm const.}\ \bb ^{6} e^{2\pi ^2/\bb }
\ \left( {\pi \mm \over \sinh \pi \mm} \right)^4 \ .}
The corresponding density of states as a function of the energy
should then have again the same leading behavior as in flat space
$d_\E\sim e^{4\pi \sqrt{\E }}$.

%%%%%%%%%%%%%%%%%%%%%%
\subsec{Model with NS-NS 3-form background }
%%%%%%%%%%%%%%%%%%%%%%

Let us now compute the asymptotic density of states in the NS-NS
model \zs\ with the spectrum  \nlic,\nklic. The
corresponding type II superstring partition function \qual\ is
given by the product of
the  equal left $Z_L$ and right $Z_R$ sector
contributions with\foot{In contrast to the R-R model, here we will
keep the parameters $f_1$ and $f_2$ (i.e. $\mm_1$ and $\mm_2$)
general. As in section 4 we use hats to indicate the quantum
operators.}
$$
Z_R(\bb , \mm_1, \mm_2)={\rm Tr}\ \big[ e^{-\bb (\hat
N_R-\mm_1\hat J_{1R} -\mm_2 \hat J_{2R})}\big] $$ \eqn\parpar{
=\prod_{n=1}^\infty {\big[1+2e^{-\beta n} \cosh(\ha \beta
(\mm_1+\mm_2) )+ e^{-2\beta n}\big] ^2 \big[1+2e^{-\beta n}
\cosh(\ha \beta (\mm_1-\mm_2) )+ e^{-2\beta n}\big]^2 \over \big(
1-e^{-\beta n} \big)^4 [1-2e^{-\beta n} \cosh(\beta \mm_1 )+
e^{-2\beta n}] [1-2e^{-\beta n} \cosh(\beta \mm_2 )+ e^{-2\beta
n}]} } 
We  used the  GS formalism expressions for the angular
momentum operators \eig , where the spin operators are
$$
s_{1R} =  \sum_{n=1}^\infty \big( b^\dagger_{1n+}  b_{1n+}  -
b^\dagger_{1n-}  b_{1n-})-{i\over 4} S_0\gamma^{12}S_0 -{i\over 2}
\sum_{n=1}^\infty S_n^\dagger \gamma^{12}S_n \ ,
$$
$$
 s_{1L} =  \sum_{n=1}^\infty \big( \tilde b^\dagger_{1n+} \tilde b_{1n+}  -
\tilde b^\dagger_{1n-} \tilde b_{1n-})-{i\over 4} \tilde
S_0\gamma^{12} \tilde S_0 -{i\over 2} \sum_{n=1}^\infty \tilde
S_n^\dagger \gamma^{12} \tilde S_n\ .
$$
 $s_{2R,L}$ are given by similar expressions 
 with   $b_{1n}\to b_{2 n}$ and
$\gamma^{12}\to \gamma^{34}$. Here
 $b_n,b_n^\dagger $ are the bosonic mode operators in the planes 12 and 34
satisfying $[b,b^\dagger]=1$.
 The fermion mode operators satisfy
$\{ S_n^{a\dagger}, S_n^b\}=\delta^{ab}$, $\{S_0^a, S_0^b\}
=\delta^{ab}$. 
The expression  \parpar\ can be  written also  as \eqn\doss{
 Z_R(\bb , \mm_1, \mm_2)=
{ e^{{\bb\over 4}} \sinh(\ha \bb \mm _1)\sinh(\ha \bb \mm _2)
\theta_2^2 (\ha \tau  (\mm_1+\mm_2) ,\tau ) \theta_2^2 (\ha \tau
(\mm_1-\mm_2) ,\tau )
 \over 4\ \k^6(\b )\cosh ({1\over 4}\beta (\mm_1+\mm_2) )
\cosh ({1\over 4}\beta (\mm_1-\mm_2) ) \theta_1(\tau \mm_1 ,\tau )
\theta_1( \tau \mm_2 ,\tau ) }\ , } where $ \tau={i\bb \over 2\pi
}$, $ \ \k\equiv \prod_{n=1}^\infty (1-e^{-\bb n})\ \ $ 
 and $\theta_{1,2}(z,\tau )
$ are  the Jacobi theta functions. As in \zrig, here we have omitted
the zero mode factor (associated with the Landau quantum numbers
in \eig\ and the fermion zero modes). Including the zero-mode
contribution
 modifies
\doss\ by an extra factor
which cancels the hyperbolic sine and cosine functions in
\doss . Using the property $\theta_1(z+\tau ,\tau )= -e^{-i\pi
\tau -2i \pi  z}\theta_1(z ,\tau ) $, we see that the partition
function  (with zero mode contribution included) is periodic in
$\mm_1, \mm_2 $ with period 2, i.e.  is invariant under $\mm_1\to
\mm_1+2n_1,\ \mm_2\to \mm_2+2n_2 $, with integer  $n_1,n_2$.
 This is a manifestation of  the periodicity of the full
string theory under such shifts of the magnetic field parameters
discussed  at the end of  section
 4 (see also \refs{\tserus,\rshet }).
In the special case $\mm_1=\mm_2 $ the period of $\mm$ is 1.

Closely related computation (with $\mm_2=0$) in the bosonic case
was done in \russu,
where, however, a magnetic field parameter was used only as a
Lagrange multiplier for the angular momentum operator, i.e. the
density of states did not depend on it.

By using the modular transformation property of the
$\theta$-function, one obtains
the following behavior at small $\bb
$:
 \eqn\pio{ Z_R(\bb ,\mm_1, \mm_2) \cong {\rm const. }\
\bb^{6} e^{2\pi^2/\bb } {\mm_1 \mm_2 \over \sin (\pi \mm_1) \sin
(\pi \mm_2) } \ .}
Interestingly, $Z_R$ \pio \ formally coincides (after setting
$\mm_1=\mm_2 $ in \pio) with the partition function \zzzjj\ of the
R-R spectrum, provided we make the substitution $\mm\to i \mm $.

\medskip

The number of states with given energy $ \E $ and given angular
momenta $ J_{1R},J_{2R} $ can be computed as follows. Let us make
the following replacement in \pio: \ $\mm_1\to \mm_1 +i {k_1\over
\bb}$, $\mm_2\to \mm_2 +i {k_2\over \bb}$, and then consider the
expansion \eqn\zzjj { Z_R(\bb , \mm_1+i{k_1\over \bb }, \mm_2+i{
k_2\over \bb} )=\sum_{N_R ,J_{1R},J_{2R} } d_{N_R,J_{1R},J_{2R}}\
e^{-\bb (N_R-\mm_1 J_{1R}-\mm_2 J_{2R})} e^{ik_1 J_{1R}+ik_2
J_{2R}} \ . } Then \eqn\delos { d_{N_R,J_{1R},J_{2R}}=\int d\bb \
e^{\E_R\bb }\int _{-\infty }^\infty dk_1dk_2\ e^{-i (k_1
J_{1R}+k_2 J_{2R})} Z_R(\bb , \mm_1+i{k_1\over \bb }, \mm_2+i{
k_2\over \bb} ) \ , }
$$
\E_R\equiv N_R-\mm_1 J_{1R}-\mm_2 J_{2R}\ .
$$
The integrals over $k_1$, $k_2$ can be computed exactly, by using
the identity \eqn\idden{ \int _{-\infty }^\infty dk e^{-i k J}
{k-i\mm\beta \over \sinh(\pi (k-i\mm\beta )/\bb )}= {\bb ^2e^{\bb
\mm J}\over 2\cosh^2(\bb J/2)} \ . } The remaining integral over
$\bb $ is evaluated as usual in a saddle point approximation.

Combining the left and right sectors and imposing the level
matching constraint we get, following \russu , \eqn\sya{
d_{\E,J_{1R},J_{1L},J_{2R},J_{2L} } =(d_{N_R,J_{1R},J_{2R}}\
d_{N_L,J_{1L},J_{2L}})_{ _{N_R=N_L}} \ , } where
$d_{N_R,J_{1R},J_{2R}}$ is given by ($d_{N_L,J_{1L},J_{2L}}$ has a
similar form) \eqn\drdl{ d_{N_R,J_{1R},J_{2R}}= {\rm const. }\ {
h_R^{-19/4}\over \cosh^2( {\pi J_{1R}\over \sqrt{2h_R} } )
\cosh^2( {\pi J_{2R}\over \sqrt{2h_R} } )} \exp \left[ {2\pi
(N_R+h_R)\over \sqrt{2h_R} }\right] \ , } with \eqn\zaw{ h_R\equiv
N_R-|J_{1R}| -|J_{2R}| \ . } Note that the dependence of the
density of states \sya , \drdl\ on the magnetic fields $\mm_1,
\mm_2$ is effectively hidden inside $N_R=N_L$, since \eqn\nnn{
N_R=N_L=\ha [\E+ \mm _1(J_{1R}-J_{1L})+\mm _2(J_{2R}-J_{2L}) ] \ .
} Only the sector with $J_{1R}=J_{1L}$, $J_{2R}=J_{2L}$ (or with
$J_{1R}-J_{1L}+J_{2R}-J_{2L}=0$ in the $\mm_1=\mm_2 $ case) has
density of states which is insensitive to the values of the
magnetic field parameters $\mm_{1}, \mm_2$ . In general, the
exponent in \drdl\ changes as the magnetic parameters are varied.

The total number of states with given energy can be found by
integrating \sya , \drdl\ between the Regge trajectories. The
integral seems hard to compute in a closed form, but an estimate
is made by noting that for any $|J_{1R}|,\ |J_{2R}| <
N_R^{1-\epsilon }$, the contributions of the angular momenta
$J_{1R}, J_{2R}$ can be neglected as compared to $N_R$. This holds
true in the relevant integration region, since outside this
interval the integrand is exponentially suppressed due to the
factor coming from the hyperbolic cosine in \drdl. As a result, we
find \eqn\syaa{ d_{\E } = {\rm const. }\ \E ^{-15/4} \ \exp \big(
{4 \pi \sqrt{\E }}\big) \ , } which is the same behavior as in
flat space. In the above derivation we have implicitly assumed
that $\mm_{1,2} < O(\sqrt{N_R})$. This is certainly the case: as
mentioned above, this theory is, in fact, periodic in $\mm _{1,2}$
with period 2, so that $\mm_{1,2} $ can be restricted to the
interval $0<\mm_{1,2} <2$ (in the interval $2<\mm_{1,2} <4$, the
roles of the zero-mode oscillators and some non-zero mode
oscillators are exchanged).

%%%%%%%%%%%%%%%%%%%%%%%%%%%%%%%%%%%%%%%
\newsec{Plane-wave models with reduced
supersymmetry }
%%%%%%%%%%%%%%%%%%%%%%%%%%%%%%%%%%%%%%%

A simple way to generalize the above plane wave models maintaining
their solvability but reducing the number of supersymmetries is by
adding twists in spatial 2-planes. This amounts to shifting the
polar angles of eq.~\stoi\ as $\varphi _{1,2} \to
\varphi'_{1,2}+b_{1,2} \psi$, where $\psi\equiv \psi + 2\pi R$ is
a compact coordinate of a circle of radius $R$ (see e.g.
\refs{\rrt,\super}). The corresponding background is locally (but,
for generic $b_{1,2}$, not globally) equivalent to \stoi. While in
the special case $b_1=b_2$, this procedure breaks only half of the
supersymmetry \super, for generic $b_1, b_2$ all supersymmetries
are broken. For the fermions, the effect of the twists is
equivalent to adding a locally trivial connection.

We will consider two cases: when $\psi$ is one of the extra 4 flat
coordinates $x_\I $, and the case when $\psi$ is the same as
$y=\ha (u+v)$. In the latter case we will need first to generalize
the previous discussion to the case when $y$ direction in \ett\ is
periodic.

%%%%%%%%%%%%%%%%%%%%%%%%%%%
\subsec{NS-NS model with $\psi=x_5$}
%%%%%%%%%%%%%%%%%%%%%%%%%%%

Here the metric in \cstro\ becomes ($\a=5,6,7,8$)
$$ ds^2 = dudv - (f^2_1 r^2_1 + f_2^2 r^2_2) du^2 $$
\eqn\ghil{ + \ dr^2_1 + r^2_1 ( d \vp_1 + b_1 d x_5 )^2 + dr^2_2 +
r^2_2 ( d \vp_2 + b_2 d x_5 )^2 +dx_\I^2\ . } The solution of the
corresponding generalization of the string model \cstro\ proceeds
along the same lines as in the case of the models in
\refs{\tserus,\rshet,\rrt}: the twists simply modify the boundary
conditions for the world-sheet fields.

Let $m$ and $w$ represent the Kaluza-Klein momentum and the
winding number associated with the $x_5$ direction. A shortcut way
to obtain the spectrum is by using the observation of \rrt\ that
the spectrum of the twisted theory can be obtained from the
spectrum of the untwisted one by the formal replacements:
\eqn\rbbb{ \eqalign{ m &\to m- b_1 R\hat J_1 -b_2 R\hat J_2\ , \cr
\ \ \ \hat N_R &\to \hat N_R- b_1 wR \hat J_{1R} - b_2 wR \hat
J_{2R}\ ,\cr \ \ \hat N_L &\to \hat N_L + b_1 wR \hat J_{1L} + b_2
wR \hat J_{2L}\ .\cr } } Then from \nwuno\ we find
$$ \a '(E^2-p_6^2-p_7^2-p_8^2)= 2 (\hat N_L +\hat N_R) + \a'R^{-2}
(m- b_1 R\hat J_1 -b_2 R\hat J_2)^2 +{w^2R^2\over{\a'}}
$$ $$+ \ \a'p^2_y - 2\a'
(p_y+E)[f_1( \hat J_{1R} - \hat J_{1L}) +f_2 (\hat J_{2R}-\hat
J_{2L} )]
$$
\eqn\wene{ -\ 2 b_1 wR (\hat J_{1R}-\hat J_{1L} ) - 2b_2 wR (\hat
J_{2R}-\hat J_{2L}) \ . }

This model has no supersymmetry if $b_1 \not=b_2$, and has 8
(counting real supercharges) supersymmetries preserved when
$b_1=b_2$.

%%%%%%%%%%%%%%%%%%%%%%%%%%%%
\subsec{NS-NS model with $\psi=y\equiv\ha (u+v)$ }
%%%%%%%%%%%%%%%%%%%%%%%%%%%%%%%
Here the generalization of the metric in \cstro\ is
$$
ds^2 = dudv - (f^2_1 r^2_1 + f_2^2 r^2_2) du^2 + dr^2_1 + r^2_1 [
d \vp_1 + \ha b_1 (du+dv) ]^2
$$
\eqn\ghill{ +\ dr^2_2 + r^2_2 [ d \vp_2 + \ha b_2 (du+dv) ]^2
+dx_\I^2 \ , } where $y=\ha (u+v)$ is assumed to have period $2\pi
R$. As was already stressed above, in the case of compact $y$
direction (and thus periodic $u$ and $v$) the string models \sto\
and \stro\ are no longer equivalent for generic $f_{1,2} R
$.\foot{The case $f_{1,2} R=n_{1,2}=0,1,2,...$, is special since
in this case the coordinate transformation \hhoo\ is globally
defined 
 and the two models are still equivalent.}

Let us first ignore the twists in the two 2-planes and generalize
the results of section 4 to the case of compact $y$. Following
\rshet\ (i.e. modifying the \lc gauge condition by an extra
$\s$-dependent term) one finds that for compact $y$ the spectrum
\wuno\ of the NS-NS model \sto\ develops the following dependence
on $wR$ \eqn\wdos{ E^2- p_\a^2={2\over\a' } (\hat N_L +\hat N_R)+
{m^2\over R^2}+{w^2R^2\over{ \a '}^2} - 4 ({m\over R} +{wR\over
\a' }+E)(f_1\hat J_{1R}+f_2\hat J_{2R} ) \ . } The spectrum
\nwuno\ of the model \stro\ is replaced in a similar way by
$$
{E'}^2-p_\a^2={2\over\a'} (\hat N_L +\hat N_R)+ {{m'}^2\over
R^2}-2 ({m'\over R}+E')\big[f_1 (\hat J_{1R}-\hat J_{1L}) + f_2
(\hat J_{2R}-\hat J_{2L})\big]
$$
\eqn\nwdos{ +\ {w^2R^2\over{\a'}^2}-\ 2 {wR\over\a '} \big[f_1
(\hat J_{1R}+\hat J_{1L}) + f_2 (\hat J_{2R}+\hat J_{2L})\big] \ .
} Here the integers $m,w$ represent the momentum and winding
numbers in the direction $y$. The spectrum \wdos\ is the extension
of the spectrum of the model studied in \rshet\ to the
two-parameter ($f_1, f_2 $) case.

The difference between the two models is due to the fact that the
coordinate redefinition in \hhoo\ now introduces a modification in
the boundary conditions. The compact version of the model
\sto,\fef\ with the spectrum \wdos\ has 12 supersymmetries if $f_1
=\pm f_2$, and none in the opposite case (see discussion below
\oup). The compact version of the model model \stro,\oup\ with the
spectrum \nwdos\ has no supersymmetries for generic $f_1,f_2$ and
8 supersymmetries if $f_1 =\pm f_2$.
 For special values $f_1R=f_2R=
n=0,1,2,...$ there is an enhancement of supersymmetry from 8 to 16
supercharges.

\bigskip

Let us now include the twists $b_1,b_2$ in the two polar angles.
Let us first consider the spectrum of the $(b_1,b_2)$ twisted
version of the model \sto,\fef. Using the prescription \rbbb\ in
the spectrum \wdos , we get
$$\a 'E^2- \a'p_\a^2= 2 (\hat N_L +\hat N_R)+
{\a'\ov R^{2}} (m- b_1 R\hat J_1 -b_2 R\hat J_2)^2
+{w^2R^2\over{\a'}}
$$
$$ - \ 4\a'
({m\over R} - b_1 \hat J_1 -b_2 \hat J_2+ {wR\over \a'} +E ) (f_1
\hat J_{1R} +f_2 \hat J_{2R})
$$
\eqn\wepe{ -\ 2 b_1 wR (\hat J_{1R}- \hat J_{1L}) - 2b_2 wR (\hat
J_{2R}- \hat J_{2L})\ \ . } The spectrum of the twisted version of
the model \stro,\oup\ whose metric is \ghill \ can then be found
using the observation that the corresponding generalizations of
the models \zs\ and \stoi\ are formally related by: (i) the shift
of the twist parameters: $b_1\to b_1+f_1 , \ b_2\to b_2+f_2$, and
(ii) the shift of the energy in \eeee . Indeed, going from \sto\
to \stro\ is equivalent (see \hhoo) to the twist $\varphi_{1,2}\to
\varphi_{1,2} + f_{1,2}(y-{\rm t})$. The shift by $f_{1,2} {\rm
t}$ produces the redefinition of the energy \eeee, while the shift
by $f_{1,2} y$ is equivalent to the redefinition of the parameters
$b_{1,2}\to b_{1,2}-f_{1,2}$. As a result, we find
$$ \a '( {E'}^2- p_\a^2) = 2 (\hat N_L +\hat
N_R)+ {\a'\ov R^{2}} (m'- b_1 R\hat J_1 -b_2 R\hat J_2)^2
+{w^2R^2\over{\a'}} $$
$$ -\
2\a' ({m'\over R} - b_1 \hat J_1 - b_2 \hat J_2+E' ) \big[ f_1
(\hat J_{1R}-\hat J_{1L} ) +f_2 (\hat J_{2R}-\hat J_{2L} )\big]
$$
\eqn\weme{ -\ 2 w R\big[ (b_1+f_1) \hat J_{1R}- (b_1-f_1) \hat
J_{1L} +(b_2+f_2)\hat J_{2R}- (b_2-f_2)\hat J_{2L} \big]\ \ . }
The same result can be obtained also by directly applying the
prescription \rbbb\ to the spectrum \nwdos . Note that for
$f_1=f_2=0$, the spectrum \weme\ reduces to that of
\refs{\rrt,\taka,\super}.

For general values of the parameters all supersymmetries are
broken. In the special case of $b_1= b_2$ and $f_1 = f_2$ (modulo
simultaneous change of sign of $b_2$ and $f_2$) the twisted
versions of the model \sto,\fef,\wepe\ and the model
\stro,\oup,\weme\ have 6 and 4 supersymmetries respectively.

%%%%%%%%%%%%%%%%%%%%%%%%%%%%
\subsec{R-R model with compact $y\equiv\ha (u+v)$ }
%%%%%%%%%%%%%%%%%%%%%%%%%%%%%%%

Finally, let us comment on the generalization of the R-R 3-form
model \zuup,\suup\ to the case of the compact $y= \ha (u+v)$ when
$u$ and $v$ are periodic with period $2\pi R$. In the case of the
\pw background supported by $F_5$-field similar to \suup\ the
formal solution for the 32 Killing spinors \blau\ is periodic in
$u$ with period $2 \pi f^{-1}$. That means that all \lc gauge
supersymmetries will be broken unless $ f R = n, \ n=1,2,...\
$.\foot{Without imposing the \lc gauge one still finds 16 Killing
spinors by restricting the constant spinor by the condition
$(\Gamma_{1234} + \G_{5678})\epsilon_0 =0$. This condition has no
solutions in the \lc gauge, as follows from $\G^v \epsilon_0 =0$
together with MW property. However, it is not clear  how to
realize this supersymmetry at the level of string theory of \rrm.}
In the present case of the $F_3$-background \suup\ for generic
$fR$ we will still have 8
supersymmetries preserved, corresponding to translations in the
``massless'' $S^A_{L,R}$ fermionic directions (the number of
supersymmetries will be enhanced to 16 if $fR$ is  integer).
This is indeed the same amount of supersymmetry which is present
in the ``compact'' version of the NS-NS model in the corresponding
``S-dual'' $H_3$-background discussed in section 2 and above.

The ``compact'' version of the R-R model can be solved by choosing
(as in the NS-NS case \tserus) the following generalization of the
\lc gauge: $u = 2\a' p^u \tau + 2 w R \s$, where $w=0,1,2, ...$ is
the string winding number and $p^u = { m\ov R } + E$,\
$m=0,1,2,...$ (cf. \dew). The fermionic action in the gauge
$\Gamma^u \theta^I=0$ is still quadratic and given by \bass\ with
$\del_a x^m \to \del_a u$. Assuming $\del_\pm u \not=0$ and
rescaling the fermions to absorb $\del_+ u$ and $\del_- u$
factors, the bosonic and fermionic parts of the action then take
the same form as in \lboy,\sett, but now with $\mm^2 = f^2 \del_+
u \del_- u = f^2 \big[(\a 'p^u)^2 - w^2 R^2 \big]$. The spectrum
is then again given by \hhar, now with $4 p_u p_v + p^2_\a \to
-E^2 + ({ m \ov R})^2 + {(wR)^2\ov \a'} + p^2_\a$ (cf. \weme) and
with the constraint \consu\ being $N_R+ N_R^0 -N_L-N_L^0=mw $.

%%%%%%%%%%%%%%%%%%%%%%%%%%%%%%%%%%%
\newsec{ Conclusions}
%%%%%%%%%%%%%%%%%%%%%%%%%%%%%%%%%%%%%%%

Let us summarize the main results of this paper. We investigated
the plane-wave limits of the NS-NS and R-R \ads backgrounds and
have shown that the resulting string models are explicitly
solvable.

In the NS-NS case, the \pw model is equivalent to a direct
generalization of the NW model. Its solution can be obtained as a
special case of a solution of the ``magnetic'' model in
\refs{\tserus , \rshet}.
The spectrum is supersymmetric and, in addition to the usual level
number operators $\hat N_R$, $\hat N_L$, it involves the angular
momentum operators $\hat J_{1R,1L}$, $\hat J_{2R,2L}$ associated
with rotations in the two planes 12 and 34 (i.e. shifts of the
angles $\varphi_{1,2}$ in the original \ads geometry \uno). In the
compact $u,v$ case, this model may be interpreted as describing a
constant magnetic field background in closed string theory.

In the R-R case, the string model can be solved in the light-cone
gauge in terms of free massive and massless world-sheet bosons and
fermions. We have worked out the operator quantization in detail,
determining the Hamiltonian and physical states in terms of the
creation/annihilation operators corresponding to the Fourier modes
of the fields.

We have studied, both in the R-R and NS-NS cases, the asymptotic
density of states, finding similar leading large \lc energy
behavior as in flat space. In the NS-NS model, we obtained an
explicit formula for the density of states with given energy and
angular momentum components. It can also be interpreted as a
density of states of a string in magnetic field.

We also discussed solvable plane-wave modes with reduced or
completely broken supersymmetry. The corresponding plane wave
backgrounds were constructed by shifting the polar angles,
$\varphi_s \to \varphi_s+b_s \psi $, in the same way as in the
Melvin model \rrt . This produces a ``twisted'' identification of
space-time points under $\psi\to \psi +2\pi R$. The resulting
models include examples with $0,4,6,8$ or $12$ real
supersymmetries. Their spectra contain as particular cases the
spectrum of the uniform magnetic field model of \rshet, as well as
the spectrum of the magnetic flux tube model of \rrt\ and its
supersymmetric generalizations studied in \refs{\taka, \super }.

\medskip

We have seen that the NS-NS and R-R \pw string models
corresponding to the two S-dual 3-form backgrounds have equivalent
supergravity parts of their spectra. In general, the S-duality
involves transforming the supergravity background, inverting
string coupling and interchanging fundamental strings with
D-strings, and thus it does not a priori imply any relation
between the stringy parts of the two weakly-coupled fundamental
string spectra.

There are a number of open problems and further directions. One
may also determine massless vertex operators, compute simplest
correlation functions and compare them with the corresponding
expressions in $AdS_3 \times S^3$ theory before the \pw limit (cf.
\oogm). Using the \lc gauge, one may also solve explicitly for the
open-string spectrum, determining possible D-brane configurations
and comparing with previous results, e.g., for D-branes in the NW
model \stan.

As is well known (see, e.g., \peet), the \ads background is the
near-horizon region of the NS5+NS1 or D5+D1 system. Adding
momentum (wave) along the string direction leads to an extremal
5-d black hole with regular horizon and finite entropy, which can
be reproduced by counting D-brane BPS states at weak coupling and
using a non-renormalization property of the entropy. It would be
interesting to investigate which part of this picture survives
taking the \pw limit, and, in particular, whether the knowledge of
the full string spectra for the \pw models can be useful for the
study of some aspects of the black-hole physics, thus
complementing implicit (supersymmetry-based) D-brane methods.

\bigskip

\noindent {\bf Acknowledgements}
%%%%%%%%%%%%%%%%%%%%%%%%%%%%%%%

\noindent We are grateful to R. Metsaev for useful discussions.
The work of A.A.T. was supported in part by the grants DOE
DE-FG02-91ER-40690, PPARC PPA/G/S/1998/00613, INTAS 991590
and by the Royal Society Wolfson research merit award.

\vfill\eject \listrefs
\end